\documentclass[aps,reprint,amsmath, amssymb, superscriptaddress,floatfix,showpacs]{revtex4-2}
\usepackage{graphicx}
\usepackage{tikz}
\begin{document}

\title{
Loxodromes in Open Multi-Section Lasers
}

\author{Christopher P. J.  O'Connor}
\affiliation{School of Mathematical Sciences, University College Cork, Ireland}

\author{Sebastian Wieczorek}
\affiliation{School of Mathematical Sciences, University College Cork, Ireland}

\author{Andreas Amann}
\affiliation{School of Mathematical Sciences, University College Cork, Ireland}

\begin{abstract}
We introduce a formalism to efficiently calculate lasing modes and optical power flow in multi-section lasers with open boundaries. 
The formalism is underpinned by a projection of the complex-valued electric field and its spatial derivative onto a suitably extended complex $\mathcal{Z}$-plane,
to reduce the order of the problem and simplify analysis.
In a single-section laser, we show that a laser mode is a \emph{loxodrome} on the  extended complex $\mathcal{Z}$-plane.
In a multi-section laser, we obtain loxodromes for individual sections of the laser. Then, a  multi-section mode is constructed by continuously concatenating individual loxodromes from each section using the open boundary conditions. A natural visualization of this construction is given by stereographic projection of the extended complex $\mathcal{Z}$-plane onto the Riemann sphere.
Our formalism simplifies analysis of lasing modes in open multi-section lasers and provides new insight into the mode geometry and degeneracy.
\end{abstract}


\maketitle

\section{Introduction}
\label{sec:intro}

With increasing miniaturisation in optical devices and the development of photonic integrated circuits, the problem of modelling optical modes in complex
configurations comprising of both active-medium and absorbing sections becomes prominent.  For a one-dimensional 
structure, the core of the problem is to find the solution 
 to a multi-point boundary value problem for the electromagnetic wave equation with complex coefficients, 
where  open boundary conditions complicate the situation.  While the single section case, which corresponds to the classical Fabry-Perot laser, can be solved analytically \cite{yariv2007photonics}, the case of two or more sections is considerably more difficult, but also much more interesting.  

\begin{figure}[ht]
    \centering
\includegraphics[width=\columnwidth]{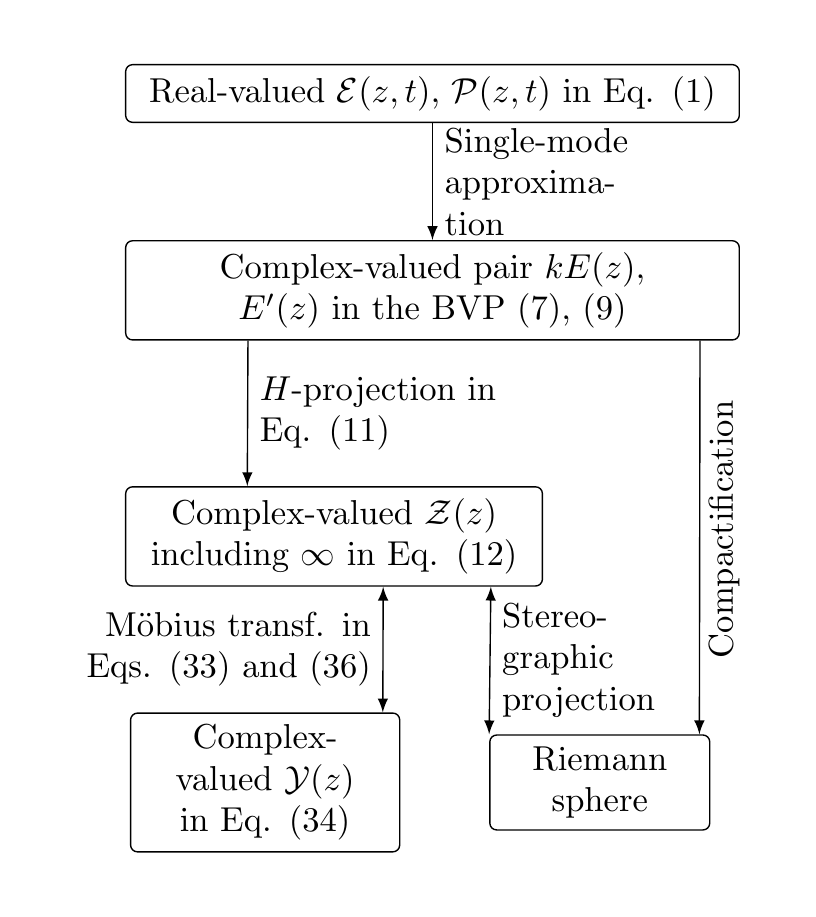}
    \caption{High level overview of our formalism, including
    different transformations involved in the three steps 
    discussed in the text.
    }
    \label{fig:steps_in_paper}
\end{figure}

The aim of this paper is to give a general method for finding lasing modes in multi-section lasers with open boundaries, and
provide a greater intuitive understanding of the geometry of lasing modes. 
To this end, we propose a formalism outlined in Fig.~\ref{fig:steps_in_paper} for laser structures in one spatial dimension denoted $z$. In the first step, 
we use the single-mode approximation to reduce the real-valued partial differential wave equation for the electric field $\mathcal{E}(z,t)$ to a complex-valued ordinary differential wave equation for the mode profiles $E(z)$.
Since the reduced wave equation is of second-order, the lasing field at each point in space is represented by two complex numbers: the electric field $kE(z)$ and its space derivative $E'(z)$. 
Hence, a lasing mode is represented by a curve in the two-dimensional complex-valued vector space (four-dimensional real-valued vector space), which  is rather difficult to visualize. In consequence, the effects of changing the pump and different laser designs are difficult to understand. 
In the second step, we address this problem of high dimensionality by a non-invertible $H$-projection of the two complex-valued variables onto a single complex-valued variable $\mathcal{Z}(z)$, with the origin of the complex $E$-plane mapped onto infinity of $\mathcal{Z}$.
The key idea is that this new variable, in conjunction with stereographic projection, provides a natural representation of a lasing mode as a one-dimensional curve on the {\em Riemann sphere}~\cite{Needham1998visual}. Note that the Riemann sphere has been used successfully in many areas of physics, for example in the guise of the Bloch Sphere representation~\cite{cohen2006quantum} of a two-level system in quantum computing, or to represent the polarization states of light on the Poincar\'e sphere~\cite{Born2013principles}. The final step of our formalism is to compute this curve on the Riemann sphere. To this end, we use the elegant mathematical formalism of the (invertible) M{\"o}bius transformation~\cite{Needham1998visual}
to show that:
\begin{itemize}
    \item[(i)]
    Each part of a lasing mode in a given section of a multi-section laser is simply a logarithmic spiral 
    on an extended complex plane.
    \item[(ii)]
    The inverse M{\"o}bius transformation of this logarithmic spiral gives a $\mathcal{Z}(z)$ that corresponds to a special curve  on the Riemann sphere called a \emph{loxodrome}~\cite{kisil2019conformal}.
    \item[(iii)]
    The entire lasing mode of a multi-section laser with open boundaries
    is obtained by concatenating 
    individual loxodromes on the Riemann sphere.
\end{itemize}

A usual approach to obtaining lasing modes in multi-section lasers is the  \emph{transfer matrix} approach~\cite{hansmann1992transfer,o2006spectral,kapon1984}.  In this approach, a part of a lasing mode in a given section of a multi-section laser is represented as a complex $2\times 2$ matrix that depends on the physical properties of this section. Our approach reduces the dimensionality of the problem from four to two real dimensions, and thus provides a simpler and more accessible visual representation of lasing modes. This makes it an interesting alternative to the transfer matrix approach.

In order to validate and demonstrate the usefulness of our approach, we first reproduce the well known results for a single-section Fabry-Perot laser. 
We then study the case of a laser with two sections of the same physical length but with different gain or absorption characteristics. In this case, we distinguish the two options where either both sections have net local gain or, alternatively, one section has net local gain while the other section has  net local absorption.  By ``local" we mean the property of the medium excluding boundaries. Finally we study a three-section laser, in which we introduce an air gap between the two outer sections with net local gain.

In this context, we focus on the interesting situations where
two different modes coalesce, or become degenerate, 
upon varying one or two system parameters.

\section{Electromagnetic Wave Equation}
\label{sec:zbasis}

The electric field inside a laser is a three-dimensional real-valued vector that varies in space and time. Consider the spatio-temporal evolution of a single (scalar) component $\mathcal{E}(z,t)$ of this field
that varies in the longitudinal $z$-direction along the laser structure~\cite{chow1999semiconductor}
\begin{align}\label{electric_field_full}
\frac{\partial^2}{\partial z^2}\mathcal{E}(z,t) -\frac{1}{c^2}\frac{\partial^2}{\partial t^2}{\mathcal{E}}(z,t)-\mu_0\frac{\partial^2}{\partial t^2}{\mathcal{P}}(z,t)=0,
\end{align}
where  $c$ is the speed of light in vacuum, $\mu_0$ is the vacuum permeability, and $\mathcal{P}(z,t)$ is the total real-valued polarisation, which is comprised of both the active medium and background polarisation components. We use a single-mode  constant-intensity approximation, and decompose the electric field and polarisation in terms of complex-valued  spatial mode profiles, denoted by $E(z)$ and $P(z)$, and temporal oscillations at an  optical
frequency $\omega$:
\begin{align}
	\mathcal{E}(z,t)&=\mathrm{Re}\left[E(z)e^{-i\omega t}\right],\\
	\mathcal{P}(z,t)&=\mathrm{Re}\left[P(z)e^{-i\omega t}\right].
\end{align}

We can now relate the same frequency components of the complex-valued polarisation and electric field~\cite{chow1999semiconductor,sargent1974m} by
\begin{align}\label{scrudu}
{P}(z)=\epsilon_0\left(\chi_b(z)+\chi_g(z)\right)E(z),
\end{align}
where $\chi_b(z)$ and $\chi_g(z)$ are the complex-valued background and 
active-medium susceptibilities, respectively.
It is useful to introduce the complex-valued permittivity of the medium $\epsilon(z)$:
\begin{align}
\epsilon(z)= 1 + \chi_b(z) + \chi_g(z).
\end{align}
The case of $\mathrm{Im}\left[\epsilon(z)\right]>0$ corresponds to 
net local absorption,
while $\mathrm{Im}\left[\epsilon(z)\right]<0$ indicates 
net local gain or absorption.
This allows us to rewrite the wave equation~\eqref{electric_field_full} in the succinct form  
\begin{align}\label{elec_field}
\left(\frac{d^2}{dz^2} + k^2\,\epsilon(z)\right)E(z)=0,
\end{align}
where 
$k=\omega/c$ 
is the free-space wavenumber. 
This second-order differential equation can be written as two coupled first-order differential equations by introducing a new variable $E'(z)$:
\begin{align}\label{e_diff_1}
\begin{split}
    \frac{d}{dz} E(z) &= E'(z),  \\ 
    \frac{d}{dz} E'(z) &= -k^2\, \epsilon(z)\, E(z).
    \end{split}
\end{align}
Since $E(z)$ and $E'(z)$ are complex-valued, we are 
dealing with a four-dimensional problem in real variables.  This is the first step shown in Fig.~\ref{fig:steps_in_paper}, in which we move from the real-valued $\mathcal{E}(z,t)$ and $\mathcal{P}(z,t)$ to the complex-valued pair $kE(z)$ and $E'(z)$.

\subsection{Boundary Conditions}
\label{sec:A}
\begin{figure}
    \centering
    \includegraphics[width=\columnwidth]{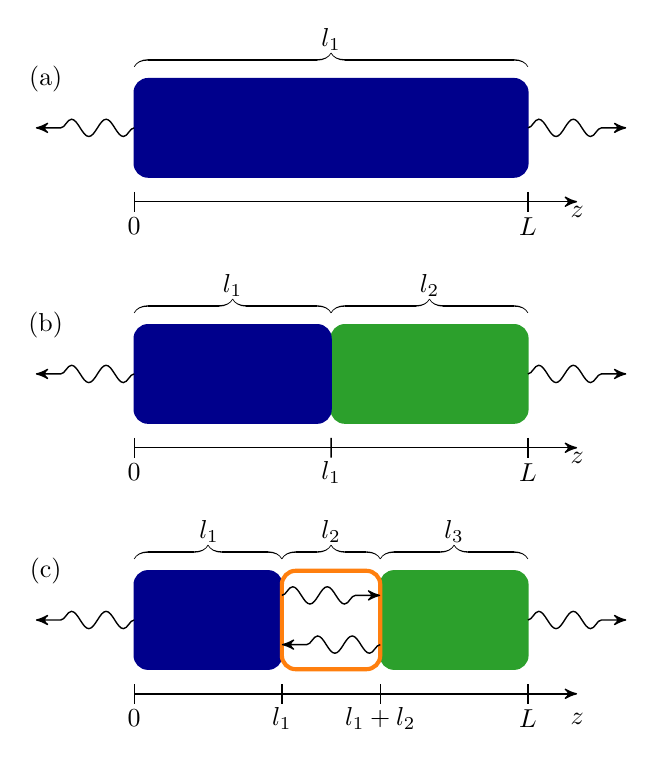}
    \caption{(a) Single-section laser with open boundaries. 
    (b) Two-section laser structure with open outer boundaries and (blue (dark grey) and green (medium grey)) two active-medium sections.
    (c) Three-section laser structure with open outer boundaries, comprising of two (blue (dark grey) and green (medium grey))  active-medium sections  separated by a (white with orange (light grey) perimeter) vacuum gap.} 
    \label{fig:setup}
\end{figure}
In this paper, we consider three different laser structures shown in Fig.~\ref{fig:setup}. The 
outer boundaries of each laser structure are at $z=0$ and $z=L$, 
and we assume only outgoing light at each outer boundary, meaning the light propagates to the left for $z<0$ and to the right for $z>L$. 
Assuming vacuum outside the laser structure, we have $\epsilon(z)=1$ for $z<0$ and $z>L$. 
Then, solving  Eq.~\eqref{elec_field} under the outgoing light assumption gives
\begin{equation}
E(z) = 
    \begin{cases}
  E(0)\,e^{-ikz} & \text{for $z<0$},\\
  E(L)\,e^{ik(z-L)} & \text{for $z>L$}.
    \end{cases}
\end{equation}
Hence we arrive at the following boundary conditions
\begin{align}\label{e_field_boundary}
\begin{split}
E'(0)&=-ikE(0),\\ E'(L)&=ikE(L),
\end{split}
\end{align}
which, together with Eqs.~\eqref{e_diff_1}, define a boundary value problem (BVP). It is important to note that this BVP does not have unique solutions: if $E(z)$ is  a solution then $\rho E(z)$ is also a solution for any complex number $\rho\neq 0$.

The $H$-projection discussed in the following section will remove this non-uniqueness.

\section{The $H$-projection}
\label{sec:hproj}
The purpose of the $H$-projection is to reduce the dimensionality of the two first-order ODEs \eqref{e_diff_1} from four real dimensions to two real dimensions.  We define the $H$-projection as a map from  $\mathbb{C}^2$ to the extended complex plane
$\hat{\mathbb{C}}=\mathbb{C}\cup\{\infty\}$ as follows:
\begin{align} 
H(h_1,h_2) = \begin{cases}
h_1/h_2 & \mathrm{for }\; h_2\neq 0,\\
\infty & \mathrm{for}\; h_2 = 0,
\end{cases}
\end{align}
where $h_1$ and $h_2$ are complex numbers.  
While $H$ is non-invertible, it removes the non-uniqueness discussed in Sec.~\ref{sec:A} in the sense that $H(\rho h_1, \rho h_2)= H(h_1,h_2)$
for any complex number $\rho\neq 0$. 
The $H$-projection corresponds to the concept of homogeneous (or projective) coordinates in the context of complex projective geometry \cite{Needham1998visual}.

Using the $H$-projection we now introduce the dimensionless function $\mathcal{Z}(z)\in \hat{\mathbb{C}}$ via
\begin{align} \label{z_basis}
\mathcal{Z}(z)= H(E'(z),kE(z)) = \begin{cases}
\frac{E'(z)}{kE(z)} & \mathrm{for }\; E(z)\neq 0,\\
\infty & \mathrm{for}\; E(z) = 0.
\end{cases}
\end{align} 
This new function allows us to to rewrite the  electric field equation \eqref{e_diff_1} and boundary conditions \eqref{e_field_boundary} as 
\begin{align}
\label{z005}
\frac{d \mathcal{Z}(z)}{dz}&=-k\left(\mathcal{Z}(z)^2+\epsilon(z)\right),\\
\mathcal{Z}(0)&=-i,
\label{z021} \\ 
\mathcal{Z}(L)&=i. 
\label{z022}
\end{align}  
This is the second step in Fig.~\ref{fig:steps_in_paper}. 

The BVP~\eqref{z005}--\eqref{z022} can be used to obtain continuous solutions on $z$-subinterval(s) where $\mathcal{Z}(z)$ is finite (or, equivalently, where $E(z)\ne0$). For example, we can choose to solve~\eqref{z005}--\eqref{z022}  where $\|\mathcal{Z}(z)\|\le 1$.
The corresponding BVP for $\mathcal{Z}(z)^{-1}$ can be derived 
as 
\begin{align}
\label{z005_inv}
\frac{d \mathcal{Z}(z)^{-1}}{dz}&=
k \left(1 + \epsilon(z) \mathcal{Z}(z)^{-2}\right),\\
\mathcal{Z}(0)^{-1} &= i,\label{z021_inv} \\ 
\mathcal{Z}(L)^{-1} &= -i, \label{z022_inv}
\end{align}  
and used to obtain continuous solutions 
on $z$-subinterval(s) where $\|\mathcal{Z}(z)\| \ge 1$, including $\mathcal{Z}(z) = \infty$ (or, equivalently, $E(z) = 0$). Then, one can match the resulting solutions at
the unit circle  $\|\mathcal{Z}(z)\| = \|\mathcal{Z}(z)^{-1}\| = 1$ to construct continuous solutions valid on the entire $z$-interval $[0,L]$.
Once a solution $\mathcal{Z}(z)$ 
is obtained, we can recover the original complex-valued electric field function $E(z)$ for a given $E(0)$ by integrating Eq.~\eqref{z_basis} to obtain
\begin{align}
\label{eq:E_field}
    E(z) = E(0) \exp\left( k\int_0^z \mathcal{Z}(z')dz' \right).
\end{align}

Since switching between $\mathcal{Z}(z)$ and $\mathcal{Z}(z)^{-1}$  is cumbersome, we propose the Riemann sphere in the next section as a more elegant way of representing solutions to the BVP \eqref{e_diff_1} and \eqref{e_field_boundary}.

\subsection{The Riemann sphere}
\begin{figure}
    \centering
    \includegraphics[width=\columnwidth]{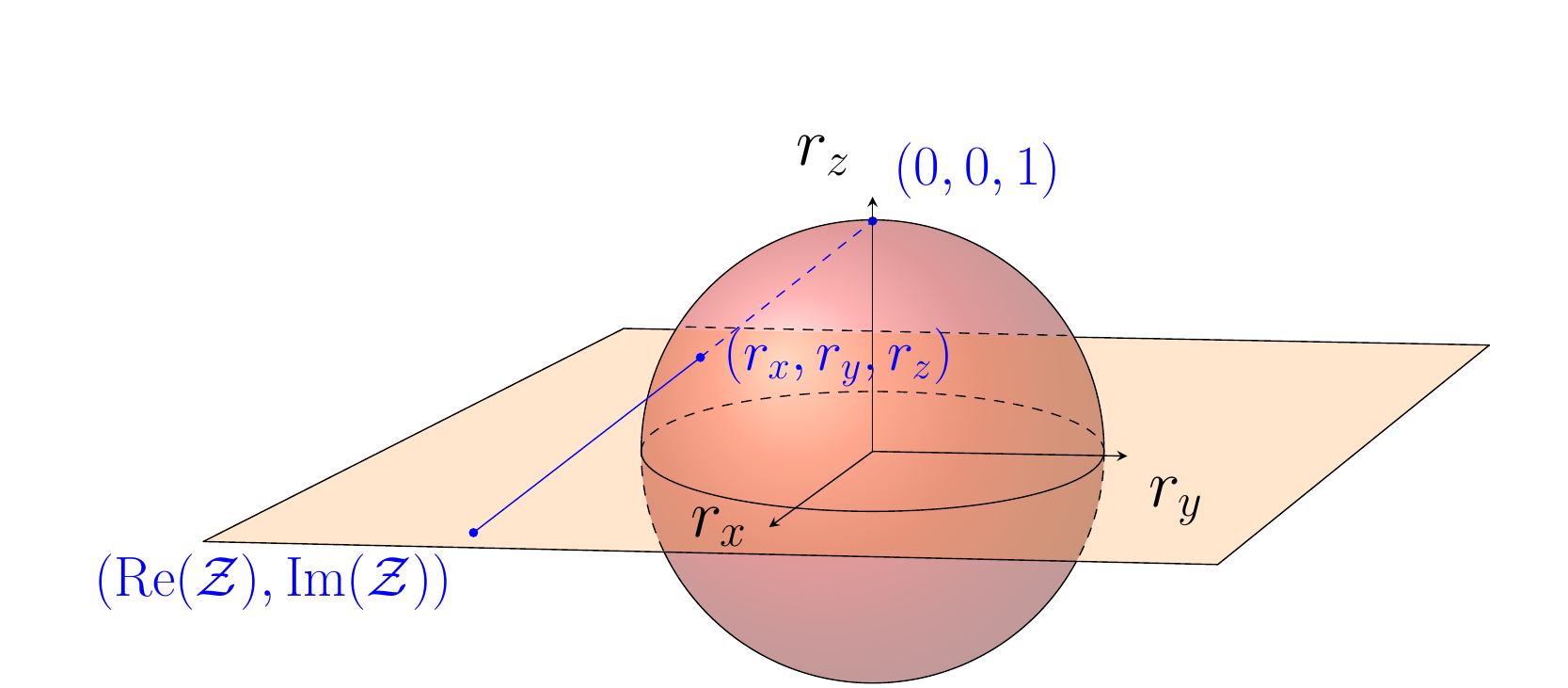}
    \caption{Stereographic projection of a point in the extended complex plane $\hat{\mathbb{C}}$
    onto the Riemann sphere embedded in $\mathbb{R}^3$ according to equation \eqref{projecting}.}
    \label{fig:stereographic}
\end{figure}
The dimensionality reduction from \eqref{e_diff_1} to \eqref{z005}
allows us to obtain intuitive insight into the nature of optical modes. 
A convenient way of visualising the extended complex plane 
$\hat{\mathbb{C}}$ is through the stereographic projection onto the {\em Riemann  Sphere}, which is given by 
\begin{align}\label{projecting}
    (r_x,r_y,r_z)=\frac{1}{1+|\mathcal{Z}|^2}\left( 2\mathrm{Re}\left[\mathcal{Z}\right], 2\mathrm{Im}\left[\mathcal{Z}\right],|\mathcal{Z}|^2-1\right).
\end{align}
Here, $(r_x,r_y,r_z)$ are coordinates in a three dimensional {\em embedding space}, and Eq.~\eqref{projecting} restricts to a sphere of radius 1 where $r_x^2 + r_y^2 + r_z^2 = 1$.
The point $\mathcal{Z}=0$ is mapped to the point $(0,0,-1)$, while  the complex infinity $\mathcal{Z}=\infty$ is mapped to the point $(0,0,1)$ in the embedding space. The boundary conditions
$\mathcal{Z}=-i$ and $i$ in equations \eqref{z021} and \eqref{z022} are mapped to points $(0,-1,0)$ and $(0,1,0)$  in the embedding space, respectively. 
This is part of the third and the last step in Fig.~\ref{fig:steps_in_paper}, which is illustrated in Fig.~\ref{fig:stereographic}. 
Since the Riemann sphere is a compact version of the extended complex plane, the last two steps of our formalism can be viewed as {\em compactification}.

\subsection{Connection to physical quantities}

Quantities of physical interest are the intensity  of the complex-valued electric field $|E(z)|^2$ and the power flow given by the time-averaged Poynting vector \cite[Ch.1.3]{yariv2007photonics},
\begin{align}
\mathcal{S}(z)=\text{Re} \left[\frac{i}{2\mu_0kc}E(z)(E'(z))^*\right],
\end{align}
where $(E'(z))^*$ is the complex-conjugate of $E'(z)$. 
Using Eq.~\eqref{eq:E_field}, we can express these quantities in terms of the $\mathcal{Z}$ function as follows
\begin{align}
\label{fint}
|E(z)|^2&=|E(0)|^2\,\exp\left(2 k\int^{z}_0\mathrm{Re} \left[\mathcal{Z}(z')\right]dz'\right),\\
\mathcal{S}(z) &= \frac{1}{2\mu_0c}\left|E(z)\right|^2\mathrm{Im}\left[\mathcal{Z}(z)\right].
\end{align} 
The interpretation of $\mathcal{S}(z)>0$ at a given $z$ is that energy flows in the positive $z$-direction (from left to right) at this $z$. Since $\mathcal{S}(z)>0$ implies $\text{Im}\left[ \mathcal{Z}(z) \right] >0$, such  energy flow 
is represented by
points on the eastern hemisphere of the Riemann sphere, i.e. points around $(0,1,0)$.  The opposite holds for $\mathcal{S}(z)<0$.

In the following, it is convenient to introduce and work with the dimensionless electric field 
\begin{equation}
\label{eq:normE}
\hat{E}(z) = \frac{E(z)}{E(0)},
\end{equation}
and the dimensionless Poynting vector
\begin{equation}
\hat{S}(z) = 2\mu_0 c\,\frac{\mathcal{S}(z)}{\left|E(0)\right|^2} = |\hat{E}(z)|^2\,\mathrm{Im}\left[\mathcal{Z}(z)\right],
\label{power_flow}
\end{equation}
Note from~\eqref{fint} and~\eqref{eq:normE} that
\begin{align}
|\hat{E}(z)|^2 &=\exp\left(2 k\int^{z}_0\mathrm{Re}\left[\mathcal{Z}(z')\right]dz'\right).
\end{align}

\section{Loxodromes for single-section lasers}

Before applying the $H$-projection to multi-section lasers, let us first illustrate its use and introduce the basic concepts in the context of the well-known single-section Fabry-Perot laser shown in Fig.~\ref{fig:setup}(a). 
We ignore any effects that cause spatial variation of $\epsilon(z)$  within the section, such as spatial hole-burning, and consider
the simplest case of constant permittivity $\epsilon(z)= \epsilon_c$.
We assume $\mathrm{Im}\left[\epsilon_c\right]<0$, which corresponds to the case of net local gain in the laser section.

\subsection{Fixed Point Analysis}
Using the definitions
\begin{align}
\mathcal{Z}^V_{1}(z)=-i\sqrt{\epsilon(z)}, \quad \mathcal{Z}^V_{2}(z)=i\sqrt{\epsilon(z)},
\end{align}
we can rewrite \eqref{z005} in the form  of a non-autonomous~\footnote{The system is non-autonomous owing to non-autonomous terms $\mathcal{Z}^V_{1}(z)$ and $\mathcal{Z}^V_{2}(z)$  with prescribed dependence on $z$.} ODE:
\begin{align} 
\label{eq:dZdz}
    \frac{d \mathcal{Z}(z)}{dz}&=-k\left(\mathcal{Z}(z)-\mathcal{Z}^V_{1}(z)\right)\left(\mathcal{Z}(z)-\mathcal{Z}^V_{2}(z)\right),
\end{align}
which holds for any spatially-varying $\epsilon(z)$.
In the special case of 
a spatially-constant $\epsilon(z) =\epsilon_c$, 
Eq.~\eqref{eq:dZdz} becomes an autonomous ODE:
\begin{align} \label{z_with_fixed}
    \frac{d \mathcal{Z}(z)}{dz}&=-k(\mathcal{Z}(z)-\mathcal{Z}^F_{1})(\mathcal{Z}(z)-\mathcal{Z}^F_{2}),
\end{align}
where 
\begin{align}
\label{z026}
\mathcal{Z}^F_{1}=-i\sqrt{\epsilon_c}, \quad
\mathcal{Z}^F_{2}=i\sqrt{\epsilon_c}.
\end{align}
We can view 
Eq.~\eqref{z_with_fixed}
as a planar autonomous dynamical system that evolves over $z$, and thus use  the concepts  of phase plane and  linear stability to give a {\em qualitative} description of solutions to \eqref{z_with_fixed}.
The two points $\mathcal{Z}^F_{1}$ and $\mathcal{Z}^F_{2}$ are \emph{fixed points}. The `stability' of these fixed points is obtained from the complex-valued Jacobian $J(\mathcal{Z})$ which is given by,
\begin{align}\label{z028}
J(\mathcal{Z})=-2k\mathcal{Z}(z).
\end{align}
We have $\mathrm{Re} \left[\epsilon_c\right]>0$ and $\mathrm{Im} \left[\epsilon_c\right]<0$. 
Using the convention $\mathrm{Re}\left[\sqrt{\epsilon_c}\right]>0$ results in $\mathrm{Re}\left[\mathcal{Z}^F_{1}\right]<0$ and $\mathrm{Im}\left[\mathcal{Z}^F_{1}\right]<0$. Using \eqref{z028}
we see that 
$\mathrm{Re}\left[J(\mathcal{Z}^F_{1})\right]=-2k\mathrm{Re}\left[\mathcal{Z}^F_{1}\right]>0$ 
and 
$\mathrm{Im}\left[J(\mathcal{Z}^F_{1})\right]\ne 0$. Therefore $\mathcal{Z}^F_{1}$ acts as an unstable spiral, meaning that solutions spiral away from $\mathcal{Z}^F_{1}$ in the phase plane $\mathcal{Z}$. Similar arguments show that $\mathcal{Z}^F_{2}$ acts as a stable spiral, meaning that solutions spiral towards $\mathcal{Z}^F_{2}$ in $\mathcal{Z}$.

\subsection{Loxodrome solution}

We now show that the general solution  to \eqref{z_with_fixed} 
has a special form known as
a \emph{loxodrome} \cite{Needham1998visual,kisil2019conformal,monzon2011geometrical}. To define a loxodrome formally, the concepts of \emph{logarithmic spiral} and M{\"o}bius transformation are required. A logarithmic spiral is a curve in $\hat{\mathbb{C}}$ given by
\begin{equation} \label{logspiral}
    Q(z) = Q_0 \exp\left[ W z \right], 
\end{equation}
where $Q_0, W\in \mathbb{C}$ and $z\in \mathbb{R}$.
A M{\"o}bius transformation
is 
a function $M$ on $\hat{\mathbb{C}}$ of the form  
\begin{equation}
    M(p) = \frac{a_{11}p+a_{12}}{a_{21}p+a_{22}},
\end{equation}
where  $p\in \hat{\mathbb{C}}$ and $a_{ij}$ are complex numbers which fulfill the condition 
$$
a_{11}a_{22}-a_{21}a_{12}\neq 0.
$$
We note that every M{\"o}bius transformation has an inverse which is also a M{\"o}bius transformation.
A loxodrome is defined as a M{\"o}bius transformation of a logarithmic spiral, i.e. as $M(Q(z))$.

To obtain the general solution to \eqref{z_with_fixed}, we consider
the following M{\"o}bius transformation from $\mathcal{Z}\in \hat{\mathbb{C}}$ to $\mathcal{Y} \in \hat{\mathbb{C}}$:
\begin{align}\label{z006}
\mathcal{Y}=\frac{\mathcal{Z}-\mathcal{Z}^F_{ 1}}{\mathcal{Z}-\mathcal{Z}^F_{ 2}}=\frac{\mathcal{Z}+i\sqrt{\epsilon_c}}{\mathcal{Z}-i\sqrt{\epsilon_c}}.
\end{align} 
This transformation maps the fixed points $\mathcal{Z}_{1}^F$ and $\mathcal{Z}_{2}^F$ to $0$ and $\infty$, respectively. When applied to \eqref{z_with_fixed}, we  obtain
\begin{align}\label{y_basis}
\frac{d\mathcal{Y}(z)}{dz}= 2i\sqrt{\epsilon_c}\,k\; \mathcal{Y}(z).
\end{align}
The general solution to \eqref{y_basis} is a logarithmic spiral 
in the form of \eqref{logspiral} with $Q_0= C$ and $W = 2i\sqrt{\epsilon_c}\,k$:
\begin{align}\label{z007}
\mathcal{Y}(z)= C\,e^{2i\sqrt{\epsilon_c}\,kz },
\end{align}
where $C\in\mathbb{C}$ is an unknown constant of integration.
Applying the inverse transformation of \eqref{z006}, namely
\begin{align}
\label{z006inv}
\mathcal{Z} = i\sqrt{\epsilon_c}\;\frac{\mathcal{Y}+1}{\mathcal{Y}-1},
\end{align}
to
\eqref{z007},
we obtain the general solution for \eqref{z_with_fixed} as
\begin{align}
\label{formula_for_z}
\mathcal{Z}(z)&=\sqrt{\epsilon_c}\;
\frac{ D \cos(\sqrt{\epsilon_c}\,kz)-\sqrt{\epsilon_c}\sin(\sqrt{\epsilon_c}\,kz)}{ D \sin(\sqrt{\epsilon_c}\,kz)+\sqrt{\epsilon_c}\cos(\sqrt{\epsilon_c}\,kz)},
\end{align}
where $D\in\mathbb{C}$ is an unknown constant of integration.
The general solution 
$\mathcal{Z}(z)$ to \eqref{z_with_fixed}, given  in \eqref{formula_for_z}, is 
a M\"obius transformation of a logarithmic spiral and therefore a loxodrome. 
\begin{figure}
	\begin{center}
		\includegraphics[width=\columnwidth]{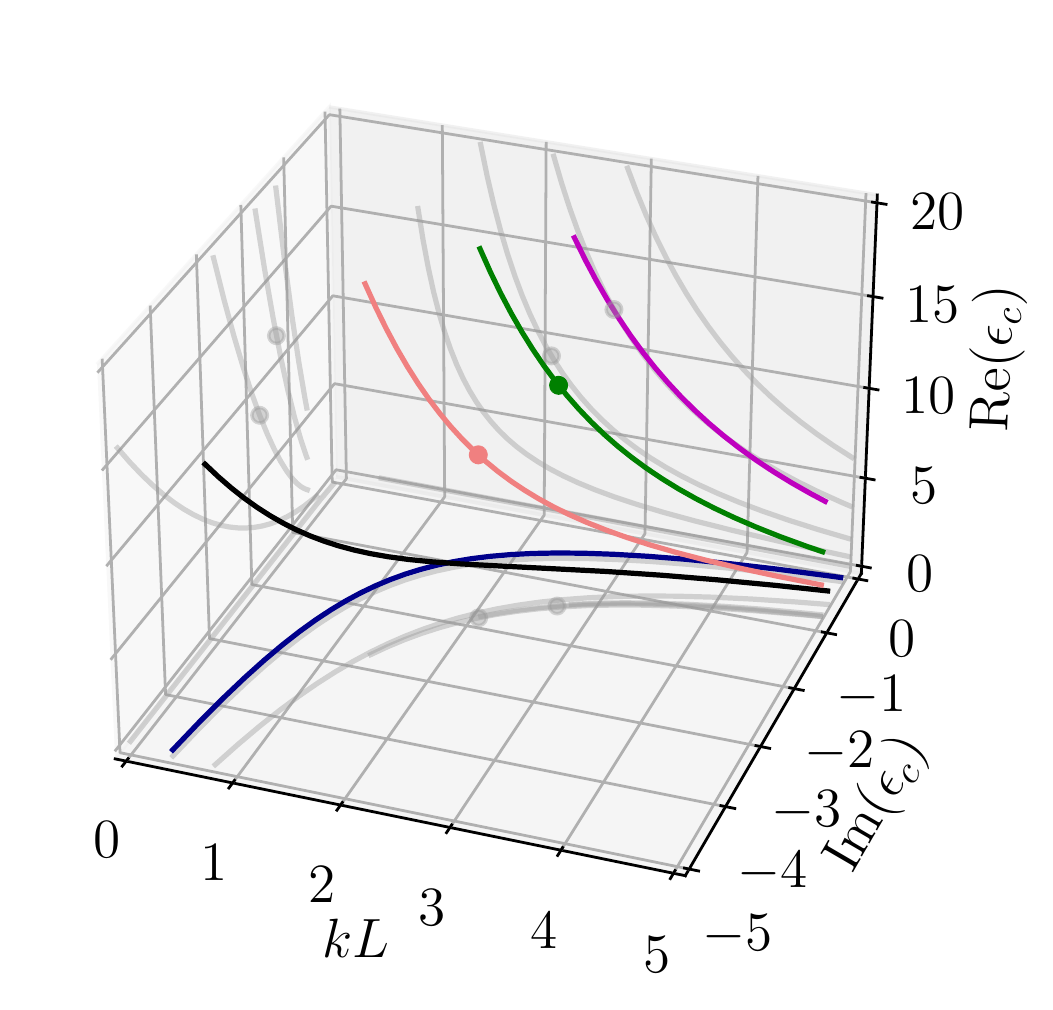}
	\end{center}
	\caption{\label{tep_2} Five solution branches of equation \eqref{z024}. The red (light grey) and green (medium grey) dots correspond to Table~\ref{parameters_table_1}.  Note that the axes in this and all following figures show dimensionless quantities.
	}
\end{figure}

\subsection{Boundary Conditions}

We now impose boundary conditions for the single-section laser to fix the unknown constant(s) of integration, and obtain combinations of $\epsilon_c$ and $kL$ that correspond to the {\em lasing modes}.

Firstly, we 
note from the general logarithmic spiral solution~\eqref{z007} that $C=\mathcal{Y}(0)$, and obtain
\begin{align}
\label{solY1}
\mathcal{Y}(z) &= \mathcal{Y}(0)\,e^{2i\sqrt{\epsilon_c}\,kz },\\
\label{solY2}
\mathcal{Y}(L) &= \mathcal{Y}(0)\,e^{2i\sqrt{\epsilon_c}\,kL }.
\end{align}
In the physically relevant case of a laser, we have  $\mathrm{Re}\left[\epsilon_c\right]>0$, $\mathrm{Im}\left[\epsilon_c\right]<0$, and $kL>0$.
Thus, Eq.~\eqref{solY1} describes a logarithmic spiral starting from a given $\mathcal{Y}(0)$,
with two unknown parameters  $\epsilon_c\in\mathbb{C}$ and $kL\in\mathbb{R}$.
Condition~\eqref{solY2} fixes  $\epsilon_c$ and $kL$ 
so that the spiral connects to a given $\mathcal{Y}(L)$. In other words, 
multiple solutions to \eqref{solY2} correspond to multiple {\em single-section lasing modes}. 
Transforming the boundary conditions \eqref{z021} and \eqref{z022} using \eqref{z006}, we obtain 
\begin{align}
  \mathcal{Y}(0) = \frac{1-\sqrt{\epsilon_c}}{1+\sqrt{\epsilon_c}}\quad\mbox{and}\quad
  \mathcal{Y}(L) = \frac{1+\sqrt{\epsilon_c}}{1-\sqrt{\epsilon_c}},
\end{align}
 and rewrite \eqref{solY2} as 
\begin{align}
\label{new}
\pm 1 &= \frac{1-\sqrt{\epsilon_c}}{1+\sqrt{\epsilon_c}}\,e^{i\sqrt{\epsilon_c}\,kL }.
\end{align}
We use this formula to illustrate lasing modes as a family of one-dimensional manifolds in the three-dimensional parameter space $(\mathrm{Re}\left[\epsilon_c\right],\mathrm{Im}\left[\epsilon_c\right],kL)$ as shown in Fig.~\ref{tep_2}. 

Alternatively, we 
note from the general loxodrome solution \eqref{formula_for_z} that $D=\mathcal{Z}(0)$, and obtain
\begin{align}
\label{solZ}
\mathcal{Z}(z)&=\sqrt{\epsilon_c}\;
\frac{\mathcal{Z}(0) \cos(\sqrt{\epsilon_c}\,kz)-\sqrt{\epsilon_c}\sin(\sqrt{\epsilon_c}\,kz)}{\mathcal{Z}(0) \sin(\sqrt{\epsilon_c}\,kz)+\sqrt{\epsilon_c}\cos(\sqrt{\epsilon_c}\,kz)}. 
\end{align}
Then imposing the boundary conditions
\eqref{z021} and \eqref{z022} yields the single-section lasing modes condition
\begin{align}
\label{z024}
i&=\sqrt{\epsilon_c}\frac{-i\cos(\sqrt{\epsilon_c}\,kL)-\sqrt{\epsilon_c}\sin(\sqrt{\epsilon_c}\,kL)}{-i\sin(\sqrt{\epsilon_c}\,kL)+\sqrt{\epsilon_c}\cos(\sqrt{\epsilon_c}\,kL)}.
\end{align}
Condition \eqref{z024} is equivalent to condition \eqref{new}, meaning that it fixes
$\epsilon_c\in\mathbb{C}$ and $kL\in\mathbb{R}$ so that the loxodrome solution $\mathcal{Z}(z)$ connects $\mathcal{Z}(0)=-i$
and $\mathcal{Z}(L)=i$.
This condition will be useful when we generalise the calculation of lasing modes  to
multi-section lasers.

\subsection{Loxodromes for Single-Section Lasers}

\begin{figure}
	\begin{center}
		\includegraphics[width=\columnwidth]{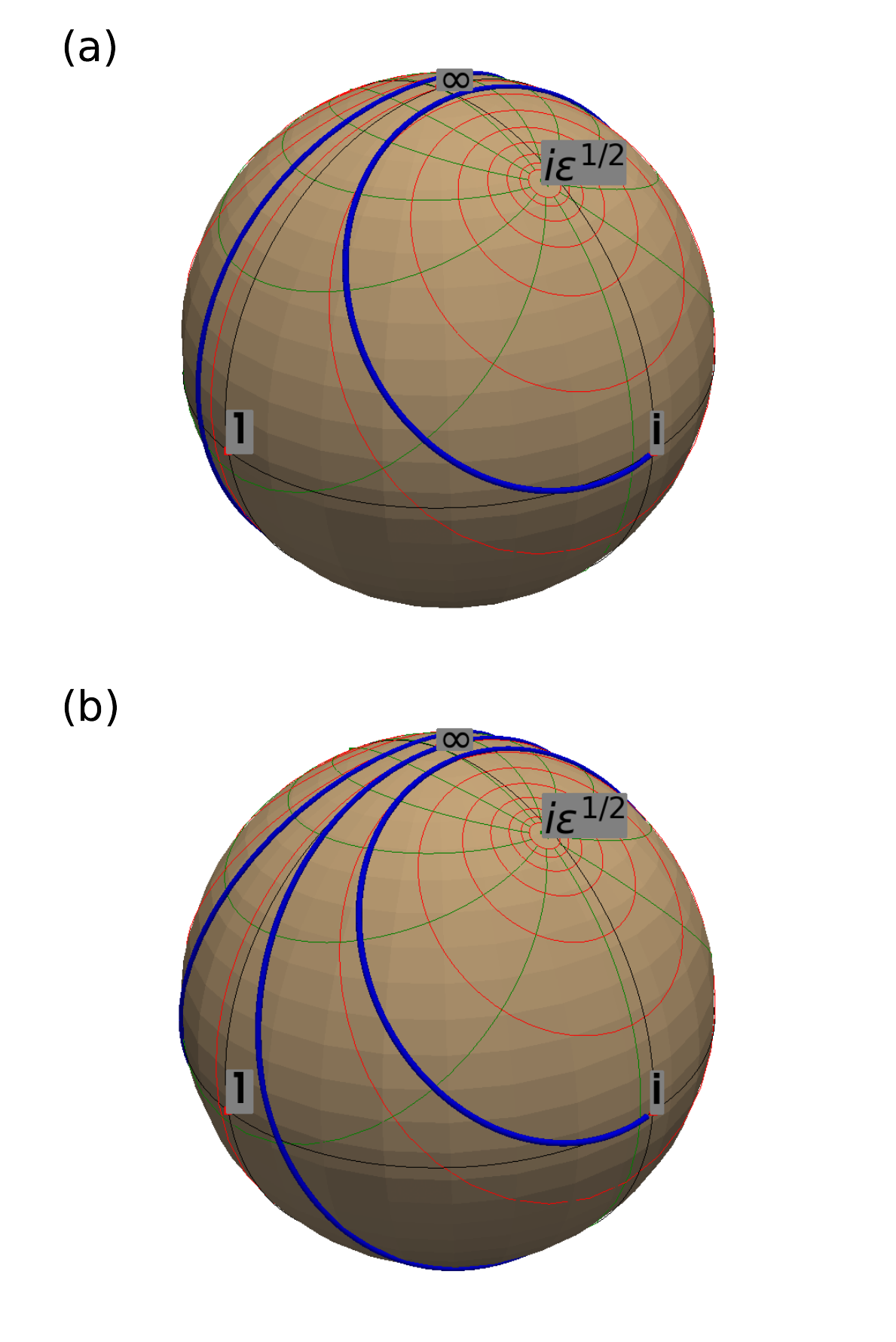}
	\end{center}
	\caption{\label{pencil} Blue (dark grey) lines show the loxodrome solutions $\mathcal{Z}(z)$ projected onto the Riemann sphere using equation \eqref{projecting}. The parameter sets for the two panels (a) and (b) are given in Table \ref{parameters_table_1}. The red (light grey) and green (medium grey) circles are the orthogonal pencils with centers $\mathcal{Z}_{1}^F=-i\sqrt{\epsilon_c}$ and $\mathcal{Z}_{2}^F=i\sqrt{\epsilon_c}$. }
\end{figure}

\begin{figure}
	\begin{center}
		\includegraphics[width=\columnwidth]{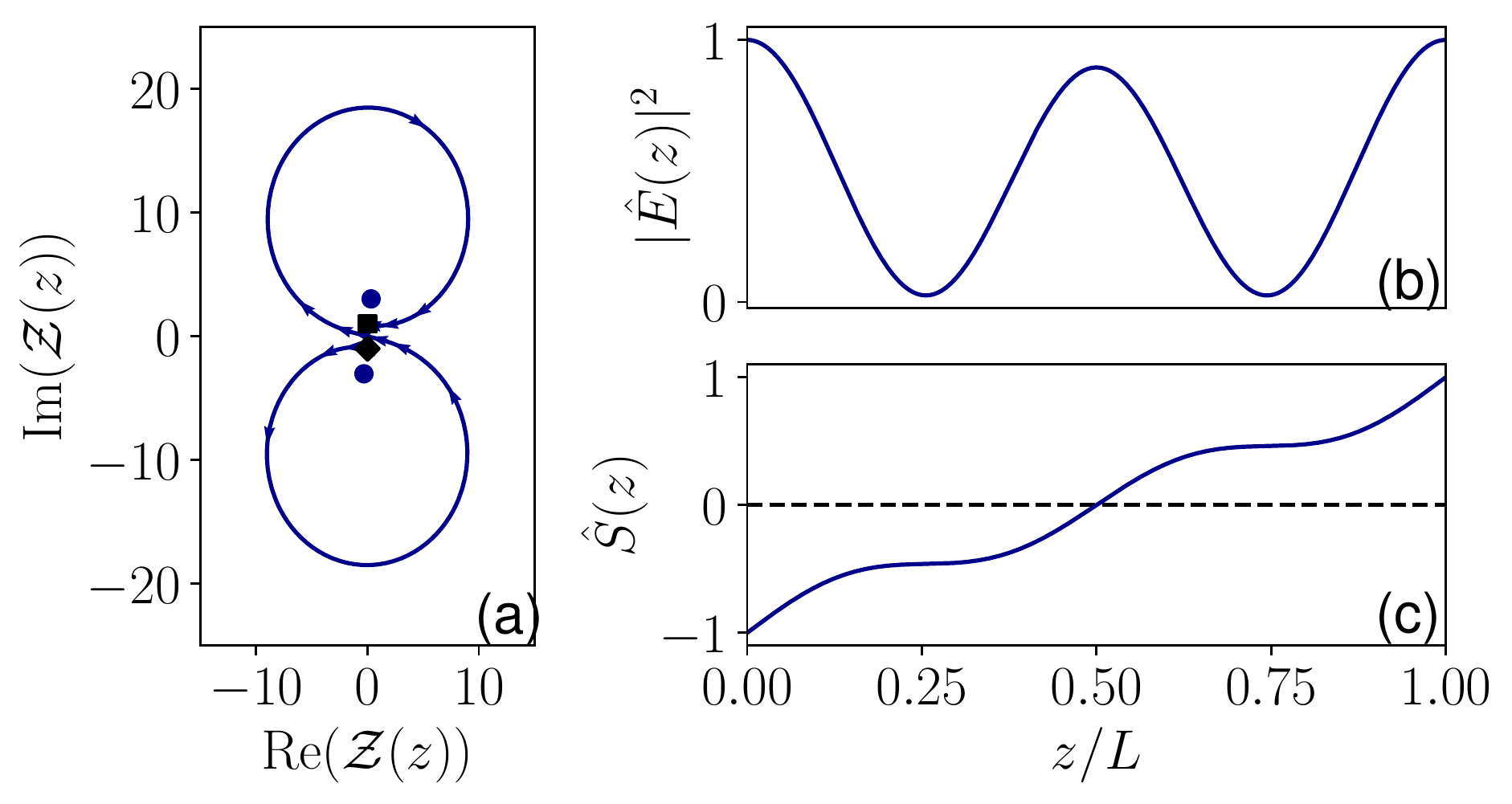}
	\end{center}
	\caption{\label{loxo_2a} (a) Parametric plot of the loxodrome solution $\mathcal{Z}(z)$ on the complex plane; the boundary  conditions  $+i$ and $-i$ are indicated by a square and a diamond, respectively; the blue (dark grey) dots indicate the   fixed points $Z_{1}^F$ and $Z_{2}^F$. Panels (b) and (c) show the corresponding electric field intensity profile $|\hat{E}(z)|^2$ and power flow profile $\hat{S}(z)$. The parameter set is given in the first column of Table~\ref{parameters_table_1}.
	}
\end{figure}

Using the tools we have introduced so far, let us now demonstrate how lasing modes in a single section laser can
be represented on the Riemann sphere and on the complex $\mathcal{Z}$ plane.  This will also allow us to connect loxodromes to physical characteristics 
such as the electric field intensity and power flow. 
Taking the parameter values corresponding to the red (light grey) dot in Fig.~\ref{tep_2} (first parameter set in Table \ref{parameters_table_1}), we obtain a solution $\mathcal{Z}(z)$ given by \eqref{solZ}. 
This solution is shown in the extended complex  plane $\mathcal{Z}$ in Fig.~\ref{loxo_2a}(a), and projected onto the Riemann  Sphere in Fig.~\ref{pencil}(a).
In Fig.~\ref{loxo_2a}(a),
we observe that the resulting loxodrome connects the boundary conditions $-i$ and $+i$ by spiralling away from  the unstable fixed point $Z_{1}^F=-i\sqrt{\epsilon_c}$, crossing through $0$, and spiraling towards the stable fixed point $Z_{2}^F= i\sqrt{\epsilon_c}$.
Equivalently, we can observe the same behaviour on the Riemann sphere in  Fig.~\ref{pencil}(a).
The electric field intensity $\left|E(z)\right|^2$ of the corresponding lasing mode can be obtained using~\eqref{fint}. From Fig.~\ref{loxo_2a}(b), we can see
the this field intensity  has three maxima and two minima inside the laser section.

\begin{figure}
	\begin{center}
		\includegraphics[width=\columnwidth]{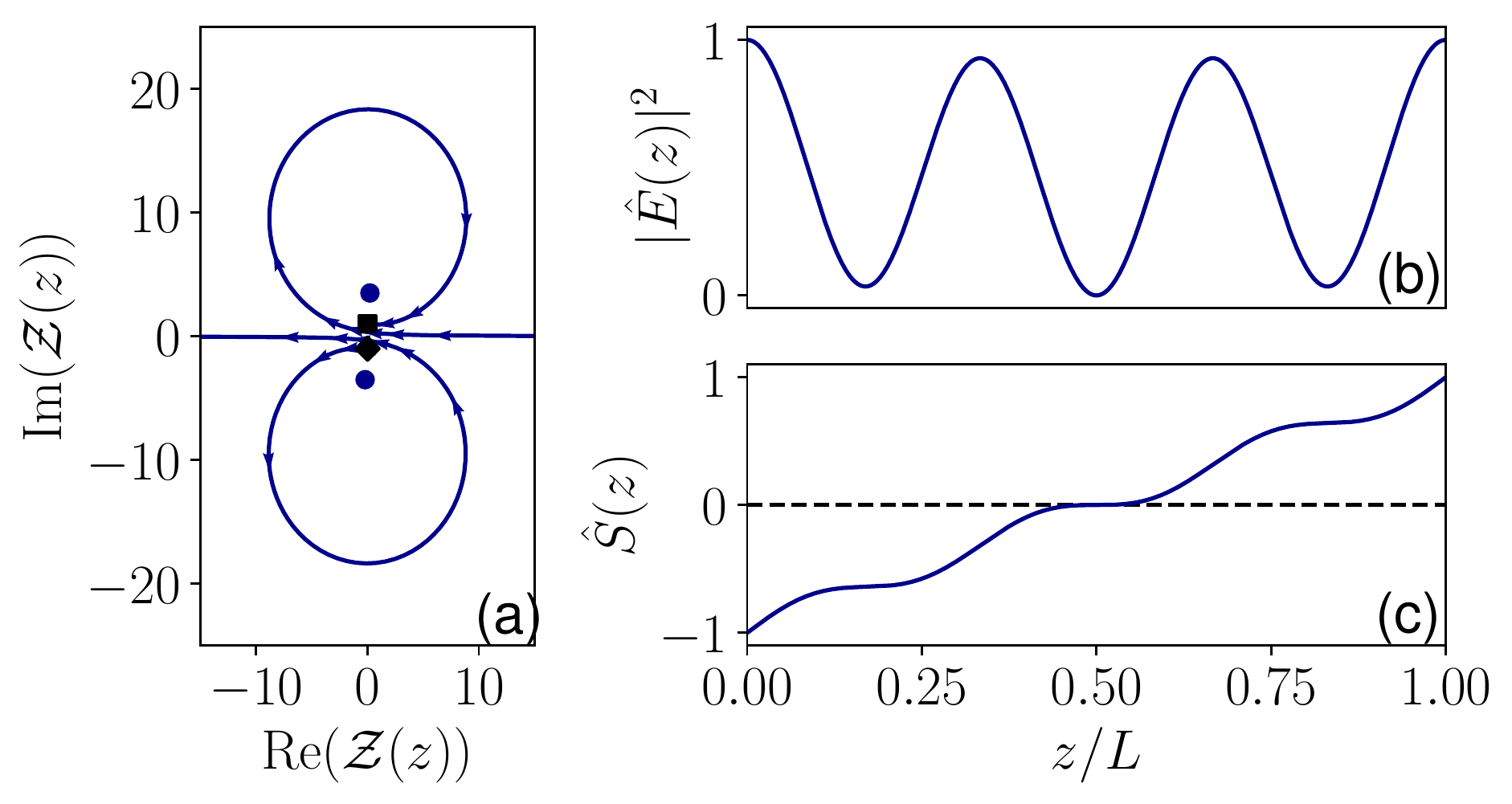}
	\end{center}
	\caption{\label{loxo_2b} As Fig.~\ref{loxo_2a} but for parameter set given in second column of Table~\ref{parameters_table_1}.
	}
\end{figure}

Similarly, taking the parameter values corresponding to the green (medium grey) dot in Fig.~\ref{tep_2} (second parameter set in Table \ref{parameters_table_1}), 
another example of a loxodrome is shown in the extended complex plane $\mathcal{Z}$ in Fig.~\ref{loxo_2b}(a), and projected onto the Riemann sphere in Fig.~\ref{pencil}(b). The key difference is that, in this instance, the loxodrome spirals through  infinity, not through 0.
The electric field intensity of the corresponding lasing mode has four maxima and three minima inside the laser, and it vanishes at the central minimum,  where  $E(L/2)=0$, or equivalently $\mathcal{Z}(L/2)=\infty$.

To provide a deeper geometrical intuition of the loxodromes on the Riemann sphere as seen in Fig.~\ref{pencil} , we include green (medium grey) circles which are representatives of the family of all circles on the sphere going through the two fixed points $\mathcal{Z}^F_{1}$ and $\mathcal{Z}^F_{2}$. In addition, we include red (light grey) circles which are representatives of the family of circles that are perpendicular to the green (medium grey) circles. Mathematically, these red (light grey) circles correspond to an \emph{orthogonal pencil of cycles} with centres $\mathcal{Z}^F_{1}$ and $\mathcal{Z}^F_{2}$ as explained in \cite{kisil2019conformal}.  The defining property of the loxodrome curve is that it crosses each family
of circles at a fixed angle   \cite{Needham1998visual}.

\begin{table}
	\begin{center}
		\begin{tabular}{ | c | c | c |}
			\hline
			\textbf{Parameters} 	& \textbf{For Figs.} 		& \textbf{For Figs.} \\
			&\textbf{\ref{pencil}(a) and \ref{loxo_2a}}&\textbf{\ref{pencil}(b) and \ref{loxo_2b}}\\
			\hline
			$kL$			& $2.1$	& $2.7$\\			
			\hline
			$\epsilon_c$& $9.0709-1.9521i$ & $12.2368-1.5171i$\\
			\hline
			$\mathcal{Z}^F_{2}=-\mathcal{Z}^F_{1}$& $0.3222+3.0290i$ & $0.2164+3.5048i$\\
			\hline 
		\end{tabular}
	\end{center}
	\caption{\label{parameters_table_1} Parameters used for Figs.~\ref{pencil}, \ref{loxo_2a} and \ref{loxo_2b}.
	}
\end{table}

Let us now connect the lasing modes of a single-section laser
to the power flow  $\hat{S}(z)$  
inside the laser
as defined in \eqref{power_flow}.  For the two examples studied above, this is shown in Figs.~\ref{loxo_2a}(c) and \ref{loxo_2b}(c), respectively. In both cases we find $\hat{S}(0)<0$ and $\hat{S}(L)>0$, which corresponds to outgoing light at either end.  In addition, $\hat{S}(z)$ increases monotonously with increasing $z$, and we have $\hat{S}(L/2)=0$.


\section{Composite Loxodromes for Multi-Section Lasers}
\label{sec:coupledcav}
Following the use of the $H$-projection for the single-section Fabry-Perot laser, our next aim is 
to obtain solutions
to the BVP
\eqref{z005}-\eqref{z022} in the case of a multi-section laser. This will
be realised through a composition of different M\"obius transformations, one for each section, in a way that is reminiscent of the transfer matrix approach \cite{davis1994transfer}. 

To be specific, we consider an $n$-section laser of the total length $L$, and use $l_j$ to denote the length of section $j$, so that $l_1+\ldots l_n =L$. We assume that permittivity $\epsilon(z)$ in Eq.~\eqref{z005} is a piecewise-constant function of $z$, and use $\epsilon_j$ to denote  constant permittivity in section $j$.
Furthermore, we use $z_j$ to denote the position of
the boundary between sections $j$ and $j+1$, with $z_0=0$ and $z_n=L$. 

\subsection{Composition of M\"{o}bius Transformations}

To make the calculation of multi-section loxodromes efficient, we introduce the following convenient notation for M\"{o}bius transformations. For a $2\times 2$ complex matrix
\begin{align}
A=
\left(
\begin{array}{cc}
a_{11}    &  a_{12}\\
a_{21}    &  a_{22}
\end{array}
\right),
\end{align}
we define the corresponding  M\"{o}bius transformation $[A]$ as follows
\begin{align}
[A](p)
=
\begin{bmatrix}
a_{11}&a_{12}\\
a_{21}&a_{22}
\end{bmatrix}
(p) 
\equiv 
\frac{a_{11}p+a_{12}}{a_{21}p+a_{22}}.
\end{align} 
Note that the representation of M\"obius transformations is not unique. In particular a matrix $cA$ defines the same M\"obius transformation as $A$ for any complex $c\neq 0$. Furthermore, we note that 
\begin{align}
\label{concatm}
\left(
[A]\circ [B]
\right) (p) 
=
[A B](p), 
\end{align}
meaning that the composition of M\"{o}bius transformations $[A]$ and $[B]$ is a M\"{o}bius transformation $[AB]$ given by the matrix product $AB$.

Using this notation, we rewrite transformation~\eqref{z006} for section $j$ in the form
\begin{align}\label{z008}
\mathcal{Y}_j(z)=\begin{bmatrix}
1&i\sqrt{\epsilon_j}\\
1&-i\sqrt{\epsilon_j}
\end{bmatrix}\left(\mathcal{Z}_j(z)\right),\quad z\in [z_{j-1},z_j].
\end{align} 
Similarly, general solution~\eqref{solY1} in $\mathcal{Y}$ for section $j$ can be written in the form
\begin{align}\label{z009}
\mathcal{Y}_j(z)=\begin{bmatrix}
e^{2i\sqrt{\epsilon_j}\,k(z-z_{j-1})}&0\\
0&1
\end{bmatrix}
\left(\mathcal{Y}_{j}(z_{j-1})\right).
\end{align}
Next, we invert~\eqref{z008}  to rewrite general solution~\eqref{solZ} in $\mathcal{Z}$ for section $j$ as a composition of M\"{o}bius transformations
\begin{align}
\label{Zlayer}
\begin{split}
\mathcal{Z}_j(z)=&
\begin{bmatrix}
i\sqrt{\epsilon_j}&i\sqrt{\epsilon_j}\\
1&-1
\end{bmatrix} \circ \\ &
\begin{bmatrix}
e^{2i\sqrt{\epsilon_j}\,k(z-z_{j-1})}&0\\
0&1
\end{bmatrix} \circ \begin{bmatrix}
1&i\sqrt{\epsilon_j}\\
1&-i\sqrt{\epsilon_j}
\end{bmatrix}
\left(\mathcal{Z}_j(z_{j-1})\right).
\end{split}
\end{align}
In this way, we obtain $n$ individual loxodromes, $\mathcal{Z}_1(z),\ldots,\mathcal{Z}_n(z)$, one for each section $j$. The electromagnetic field boundary conditions at the interface of two sections with different permittivities require continuity in the electric field and its first derivative~\cite{yariv2007photonics}
$$
E_j(z_j) = E_{j+1}(z_j) \quad\mbox{and}\quad E_j'(z_j) = E_{j+1}'(z_j).
$$
According to~\eqref{z_basis}, this translates into continuity in $\mathcal{Z}$ alone
\begin{align}
    \label{contZ}
    \mathcal{Z}_j(z_j) = \mathcal{Z}_{j+1}(z_j).
\end{align}
We then use the interior condition~\eqref{contZ} to concatenate individual loxodromes~\eqref{Zlayer} into a continuous but typically non-smooth {\em composite loxodrome} $\mathcal{Z}^{comp}(z)$. It is important to note that $\mathcal{Z}^{comp}(z)$ depends on $\mathcal{Z}_1(z_0)$ and $3n$ real parameters. These real parameters can be chosen as Re$(\epsilon_1),\ldots,$ Re$(\epsilon_n),$ Im$(\epsilon_1),\ldots,$ Im$(\epsilon_n),$ $kL$ and the $n-1$ ratios of section lengths $k l_1 : kl_2: \ldots :k l_n$. In Section \ref{sec:exceptional_points} we will consider a more convenient set of parameters based on  different physical characteristics of the individual sections.
Next, we need to ensure that such $\mathcal{Z}^{comp}(z)$ satisfies boundary conditions~\eqref{z021} and~\eqref{z022}. Thus, we impose
$\mathcal{Z}^{comp}(0)=\mathcal{Z}_1(z_0)= -i$ together with 
$\mathcal{Z}^{comp}(L) = \mathcal{Z}^{comp}(z_n) = i$, and use~\eqref{concatm} to arrive at
\begin{align}\label{transfer_n_sec}
 \begin{split}
 i=&\begin{bmatrix}
 \cos(\sqrt{\epsilon_n}\,kl_n)&-\sqrt{\epsilon_n}\sin(\sqrt{\epsilon_n}\,kl_n)\\
 \frac{\sin(\sqrt{\epsilon_n}\,kl_n)}{\sqrt{\epsilon_n}}&\cos(\sqrt{\epsilon_n}\,kl_n)
 \end{bmatrix} \circ\\
& \qquad\qquad\qquad\cdots\\
 &\circ \begin{bmatrix}
 \cos(\sqrt{\epsilon_1}\,kl_1)&-\sqrt{\epsilon_1}\sin(\sqrt{\epsilon_1}\,kl_1)\\
 \frac{\sin(\sqrt{\epsilon_1}\,kl_1)}{\sqrt{\epsilon_1}}&\cos(\sqrt{\epsilon_1}\,kl_1)
 \end{bmatrix}(-i).
 \end{split}
 \end{align}
This complex condition fixes all $3n$ real parameters to ensure that $\mathcal{Z}^{comp}(z)$ satisfies~\eqref{z021} and~\eqref{z022}.
Its multiple solutions correspond to multiple {\em multi-section lasing modes}.  

In practice, we avoid varying all $3n$ real parameters simultaneously and construct $\mathcal{Z}^{comp}(z)$ as follows. We fix the $3n$ real parameters using realistic values, start the first loxodrome from $-i$ when $z=z_0=0$ so that the first boundary condition~\eqref{z021} is satisfied, and proceed with loxodrome concatenation as described above. The result is a composite loxodrome $\mathcal{Z}^{comp}(z)$ whose endpoint $\mathcal{Z}^{comp}(L)$ lies somewhere on an extended complex plane.
Next, we want to relax as few of the $3n$ real parameters as possible to ensure that $\mathcal{Z}^{comp}(L)$ moves to the point $\mathcal{Z}=i$, so that the second boundary condition~\eqref{z022} is satisfied too.
Since $\mathcal{Z}=i$ is a single point on the extended complex plane, meaning it is of codimension-two, at least two of the $3n$ real parameters need to be varied  simultaneously to achieve $\mathcal{Z}^{comp}(L) = i$. In this way we obtain a family of composite loxodromes that solve the BVP~\eqref{z005}--\eqref{z022} with a piecewise-constant $\epsilon(z)$.  
A particular advantage of this approach is that it can be extended to any continuous spatially-varying
permittivity profile $\epsilon(z)$ by using a suitable piecewise-constant approximation of $\epsilon(z)$ with sufficiently large $n$.
Finally, the  electric field intensity and power flow of the corresponding multi-section lasing modes are obtained using~\eqref{fint} and~\eqref{power_flow}, respectively.

\subsection{Two-section Laser}

\begin{figure}
	\includegraphics[width=\columnwidth]{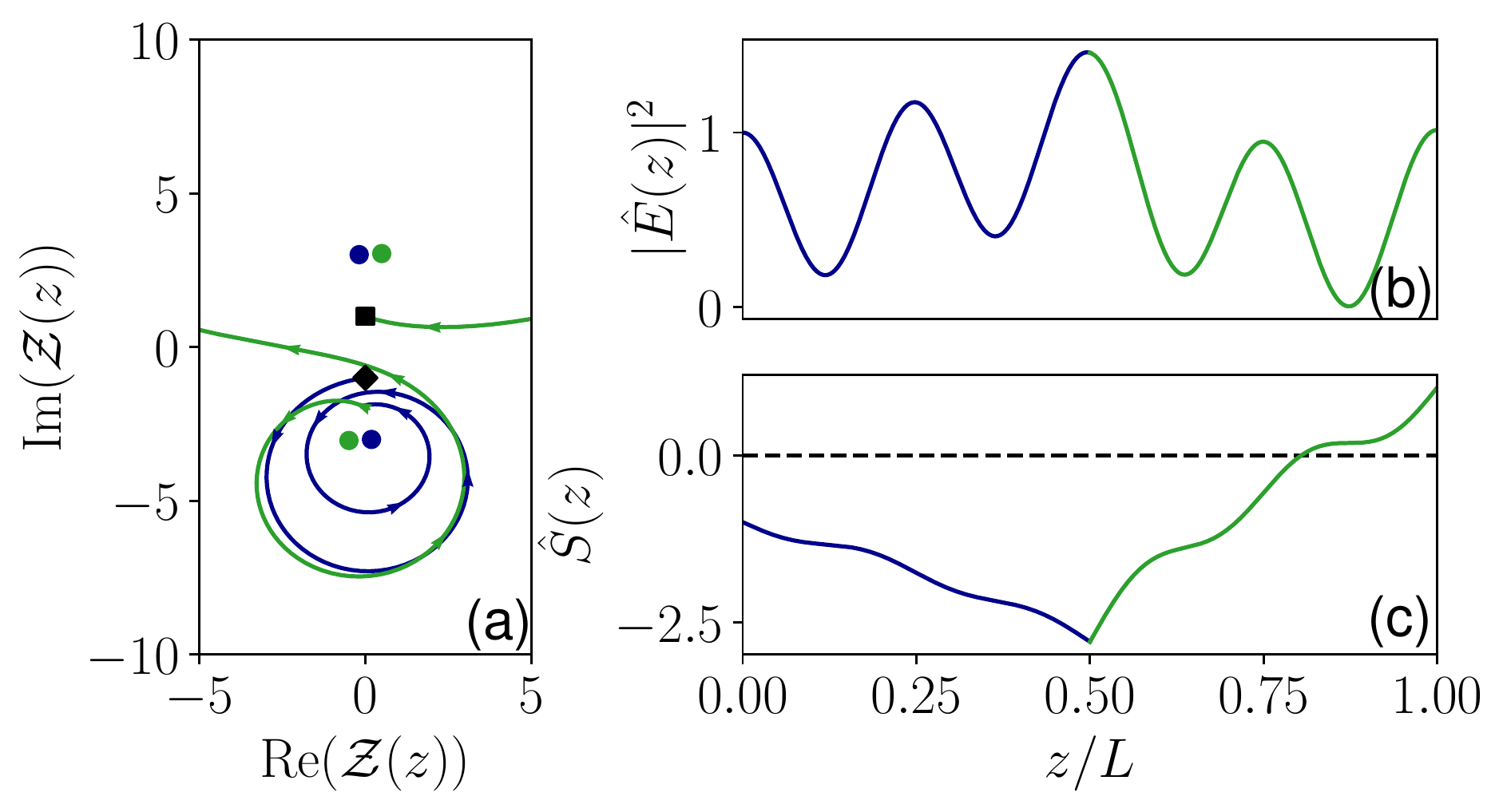}
	\caption{\label{fig:no_gap} (a) Parametric plot of the loxodrome solution $\mathcal{Z}(z)$ on the complex plane; the boundary values of $+i$ and $-i$ are indicated by a square and a diamond, respectively; the blue (dark grey) and green (medium grey) dots indicate the fixed points for sections 1 and 2 respectively. Panels (b) and (c) show the corresponding electric field intensity profile $|\hat{E}(z)|^2$ and power flow profile $\hat{S}(z)$. Blue (dark grey) curves and green (medium grey) curves denote $\mathcal{Z}(z)$ for the first and second sections respectively. The parameter values are 
	given in the first column of Table~\ref{parameters_table_no_gap}.}
\end{figure}

\begin{figure}
	\includegraphics[width=\columnwidth]{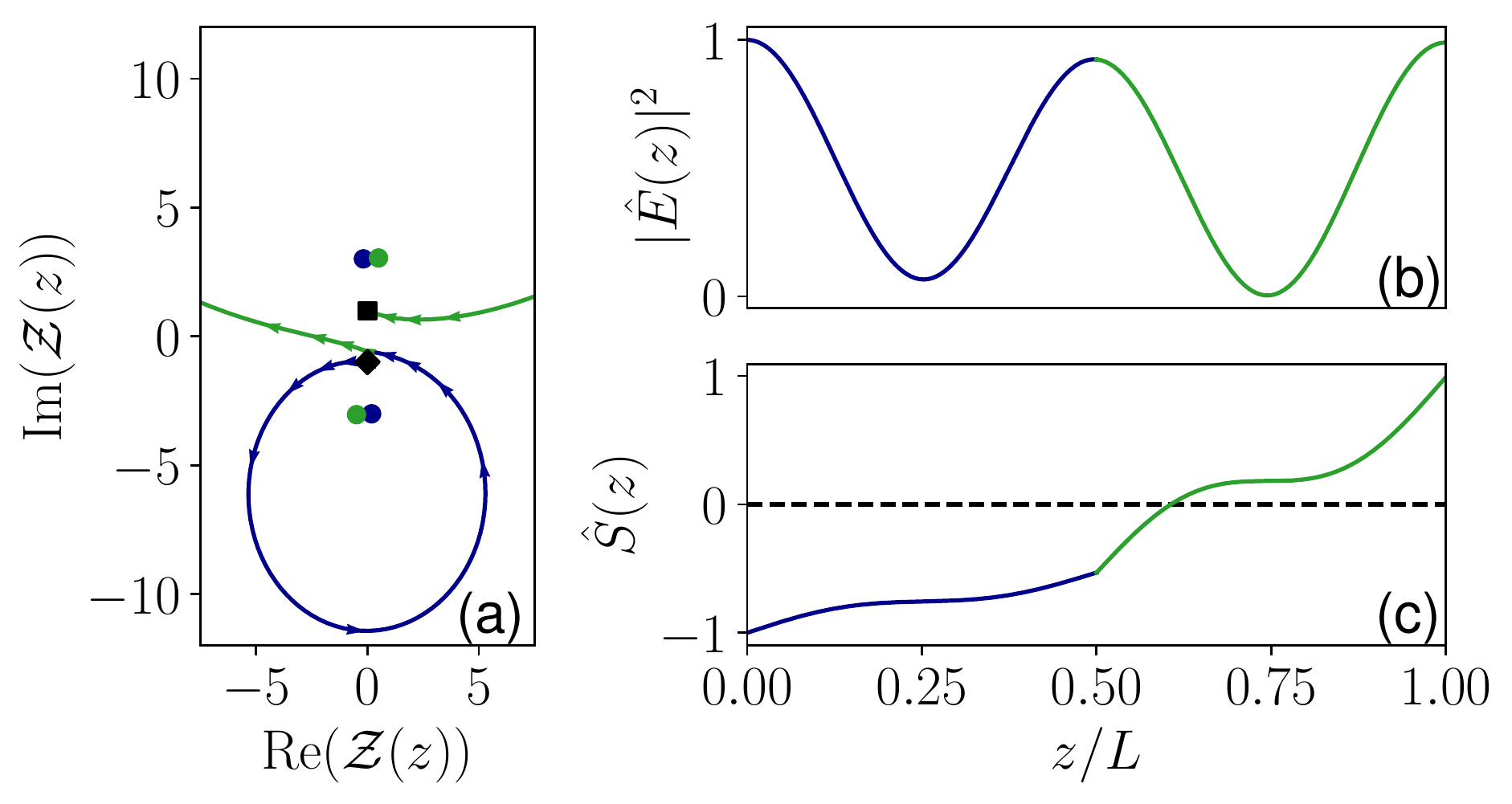}
	\caption{\label{fig:no_gap_2} The same as Fig.~\ref{fig:no_gap} but for parameter  
	values
	given in the second column of Table~\ref{parameters_table_no_gap}.}
\end{figure}

\begin{table}[t]
\begin{center}
	\begin{tabular}{ | c | c | c |}
		\hline
		\textbf{Parameters} 	& \textbf{For Fig.~\ref{fig:no_gap}} 		& \textbf{For Fig.~\ref{fig:no_gap_2}} \\
			& \textbf{(AG)}&\textbf{(GG)}\\
		\hline
		$kL$			& $4.2437$	& $2.112$\\	
    \hline
         $kl_1:kl_2$ & $1:1$ & $1:1$\\
		\hline
		$\epsilon_1$& $9.0+1.1138i$ & $9.0-0.8575i$\\
		\hline
		$\epsilon_2$& $9.0-3.0i$ & $9.0-3.0i$\\
		\hline
		$\mathcal{Z}^F_{1,2}=-\mathcal{Z}^F_{1,1}$& $-0.1853+3.006i$ & $0.1427+3.0036i$\\
		\hline
		$\mathcal{Z}^F_{2,2}=-\mathcal{Z}^F_{2,1}$& $ 0.4934+3.0402i$ & $ 0.4934+3.0402i$\\
		\hline
	\end{tabular}
\end{center}
		\caption{\label{parameters_table_no_gap} Parameters used for Figs.~\ref{fig:no_gap} and \ref{fig:no_gap_2} with the corresponding fixed points.
		}
\end{table}

Before we move on to a three-section laser, we briefly discuss a two-section laser that is characterised by six real parameters. 
A two-section laser problem has four fixed points, two for each section, which we denote $\mathcal{Z}^F_{j,1}=-i\sqrt{\epsilon_j}$ and $\mathcal{Z}^F_{j,2}=i\sqrt{\epsilon_j}$, where $j=1,2$.

First, we concatenate two loxodromes using the left-boundary condition~\eqref{z021} and the interior condition~\eqref{contZ}. Then, we vary two real parameters $kL$ and $\mathrm{Im} (\epsilon_1) $ to satisfy the right-boundary condition~\eqref{z022}. The ensuing composite loxodromes reveal two types of {\em lasing modes}: \emph{Gain-Gain} (GG) lasing modes and \emph{Absorption-Gain} (AG) lasing modes.

An example of a GG lasing mode is shown in Fig.~\ref{fig:no_gap_2} with parameter values given in Table \ref{parameters_table_no_gap}. This mode has two gain sections and is similar to the single-section lasing mode. The difference is that there are now two loxodrome parts, each with a different pair of fixed points. The loxodrome in section one (blue (dark grey)) spirals away from unstable $\mathcal{Z}^{F}_{1,1}$ towards stable $\mathcal{Z}^{F}_{1,2}$, and the loxodrome in section two (green (medium grey)) spirals away from unstable $\mathcal{Z}^{F}_{2,1}$ towards stable $\mathcal{Z}^{F}_{2,2}$.

An example of an AG lasing mode is shown in Fig.~\ref{fig:no_gap}. This mode is very different from the single-section lasing mode owing to the combination of one absorbing section ($\text{Im}(\epsilon_1)>0$) and one gain section ($\text{Im}(\epsilon_2)<0$). As a consequence, the corresponding loxodrome (blue (dark grey)) spirals towards $\mathcal{Z}^{F}_{1,1}$, which is now stable.

\subsection{Three-Section Laser}
\label{multi_without_broadening}

\begin{figure}
	\begin{center}
		\includegraphics[width=\columnwidth]{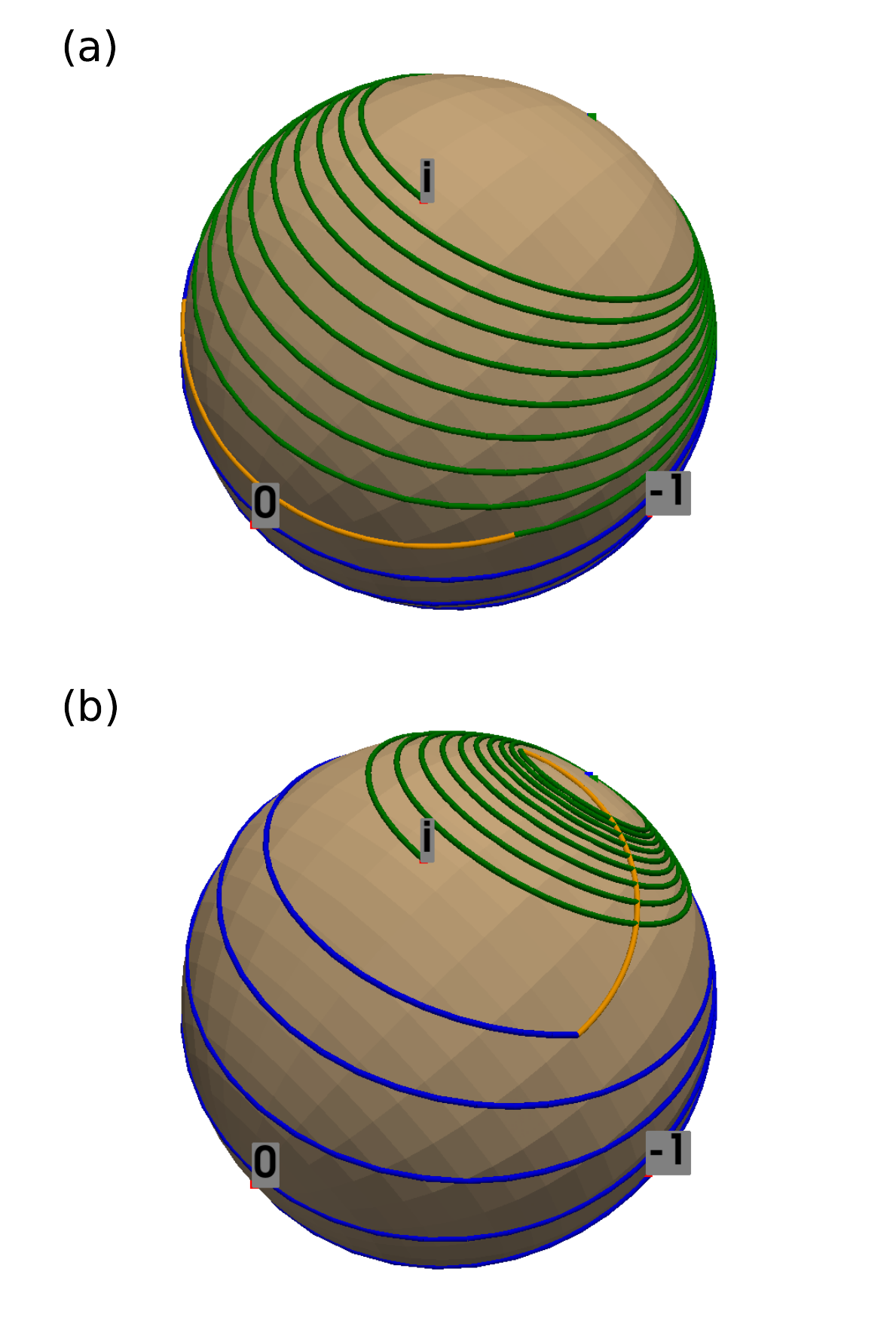}
	\end{center}
	\caption{\label{riemann_3sec} Blue (dark grey), orange (light grey) and green (medium grey) lines show the loxodrome solutions of the three section laser in the respective sections 1,2, and 3  projected onto the Riemann sphere. The parameter sets for the two panels (a) and (b) are given in Table \ref{parameters_table}.}
\end{figure}

\begin{figure}
	\begin{center}
		\includegraphics[width=\columnwidth]{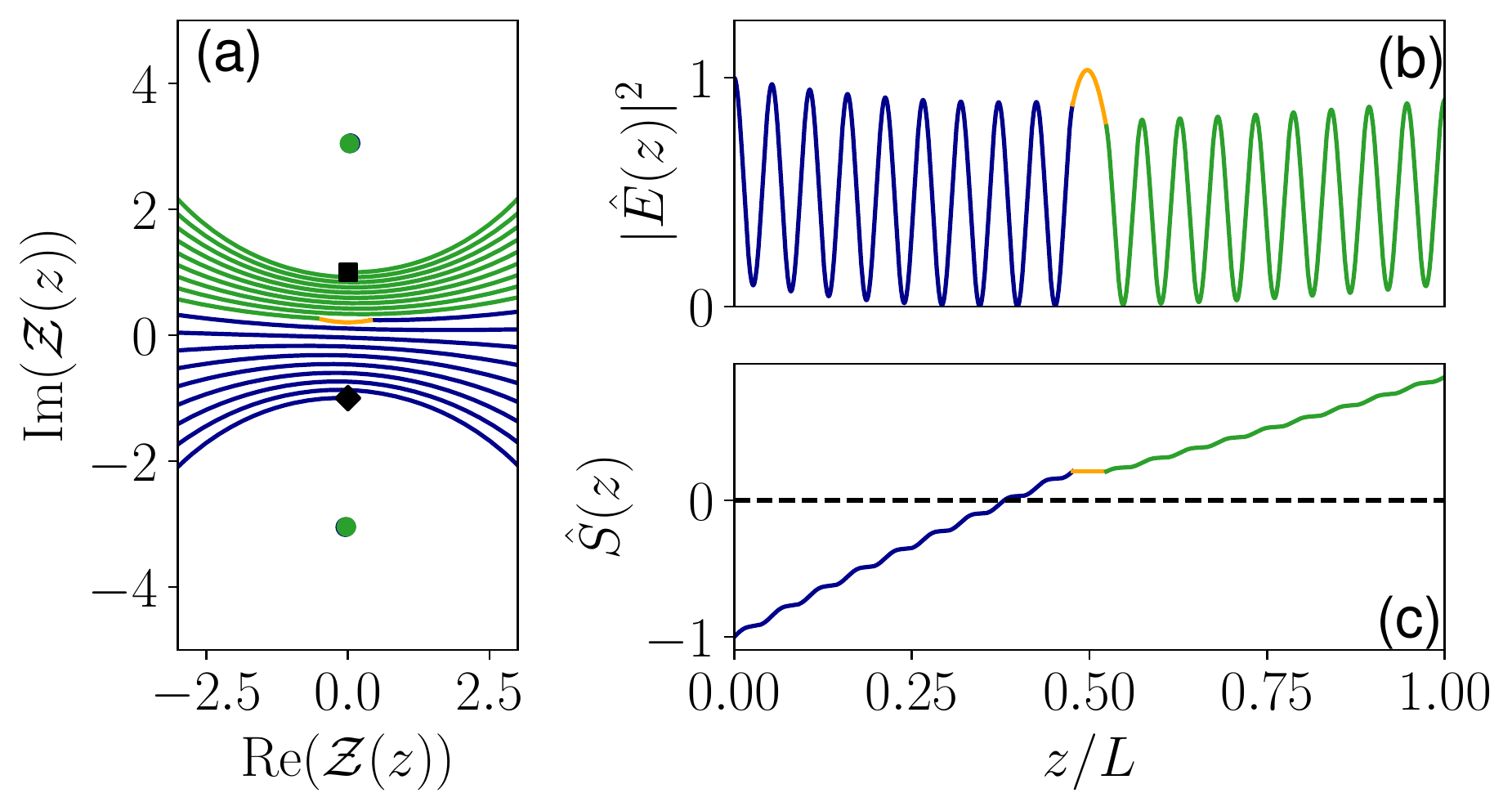}
	\end{center}
	\caption{\label{loxo_non_abs} (a) Parametric plot of the loxodrome solution $\mathcal{Z}(z)$ on the complex plane; the boundary values of $+i$ and $-i$ are indicated by a square and a diamond, respectively; the blue (dark grey) and green (medium grey) dots indicate the fixed points for sections 1 and 3 respectively. Panels (b) and (c) show the corresponding electric field intensity profile $|\hat{E}(z)|^2$ and power flow profile $\hat{S}(z)$. The colour scheme follows from Fig.~\ref{riemann_3sec} and the parameter set is given in the first column of Table~\ref{parameters_table}}
\end{figure}

\begin{table}
\begin{center}
	\begin{tabular}{ | c | c | c |}
		\hline
		\textbf{Parameters} 	& \textbf{For Figs.} 		& \textbf{For Figs.} \\
			&\textbf{\ref{riemann_3sec}(a)  and \ref{loxo_non_abs}}&\textbf{\ref{riemann_3sec}(b) and \ref{loxo_abs}} \\
			& \textbf{(GNG)}&\textbf{(GNA)}\\
		\hline
		$kL$			& $19.4456$	& $20.0942$\\	
        \hline
           $kl_1:kl_2:kl_3$ & $10:1:10$ & $10:1:10$ \\
		
		\hline
		$\epsilon_1$& $9.2503-0.3042i$ & $8.8759-0.3599i$\\
		\hline
		$\epsilon_2$& $1$ & $1$\\
		\hline
		$\epsilon_3$& $9.2086-0.1350i$ & $8.9344+0.2625i$\\
		\hline
		$\mathcal{Z}^F_{12}=-\mathcal{Z}^F_{11}$& $0.049995+3.0418i$ & $0.06039+2.9799i$\\
		\hline
		$\mathcal{Z}^F_{22}=-\mathcal{Z}^F_{21}$& $ i$ & $ i$\\
		\hline
		$\mathcal{Z}^F_{32}=-\mathcal{Z}^F_{31}$& $0.0223+3.0347i$ & $-0.0439+2.9894i$\\
		\hline
	\end{tabular}
\end{center}
		\caption{\label{parameters_table} Parameters used for Figs.~\ref{loxo_non_abs} and \ref{loxo_abs}.}
\end{table}

\begin{figure}
	\begin{center}
		\includegraphics[width=\columnwidth]{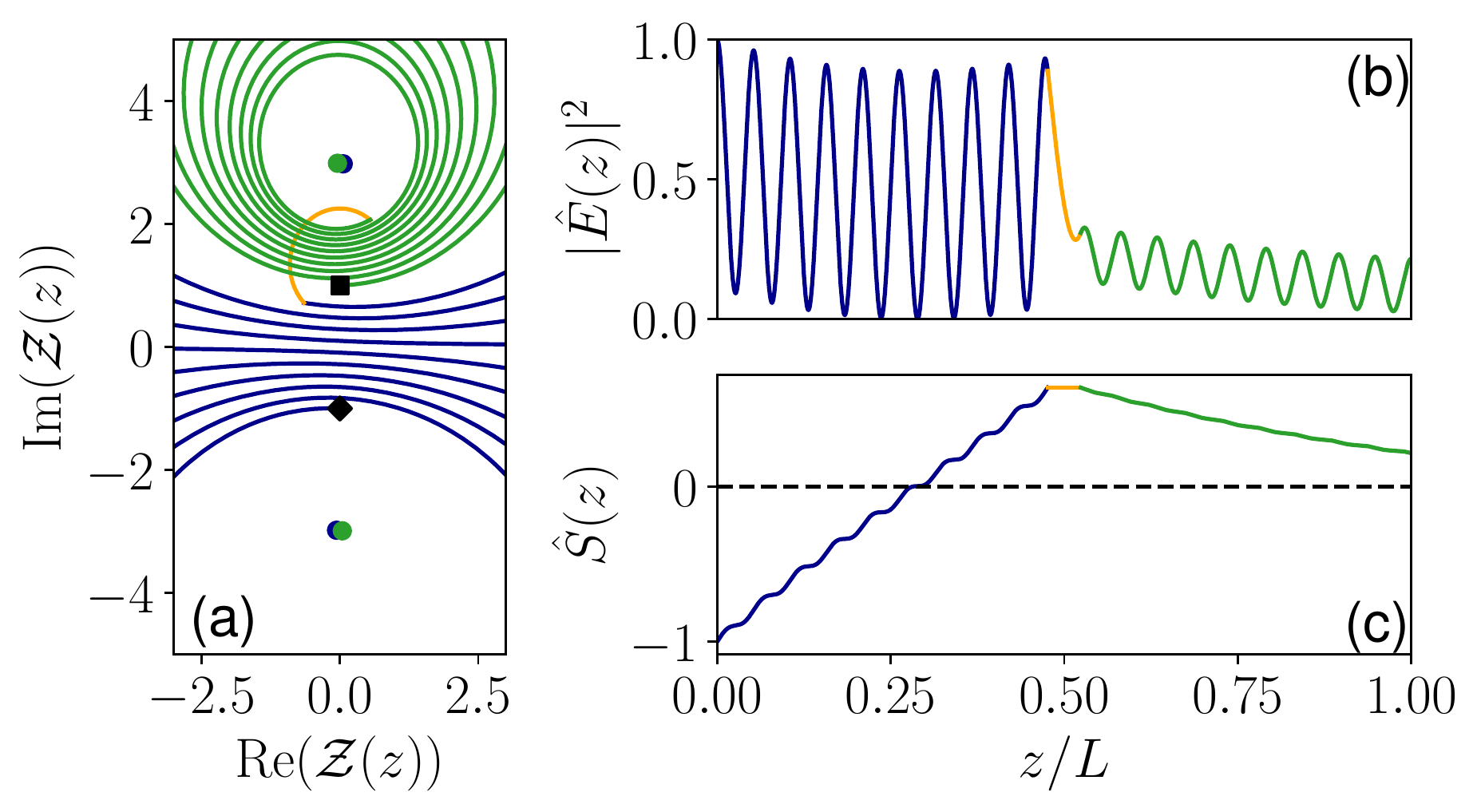}
	\end{center}
	\caption{\label{loxo_abs}As Fig.~\ref{loxo_non_abs} but for parameter set given in second column of Table~\ref{parameters_table}.}
\end{figure}

A three-section laser is characterised by nine real parameters, and has six fixed points, two for each section, which we denote $\mathcal{Z}^F_{j,1}=-i\sqrt{\epsilon_j}$ and $\mathcal{Z}^F_{j,2}=i\sqrt{\epsilon_j}$, where $j=1,2,3$.
We now discuss the specific example of a three-section laser shown in Fig.~\ref{fig:setup}(b),  where the two outer sections with local gain or absorption are separated by a section with a vacuum gap. As a result we have $\epsilon_2=1$, and thus  equation \eqref{transfer_n_sec} becomes 

 \begin{align}\label{transfer}
 \begin{split}
 i=&\begin{bmatrix}
 \cos(\sqrt{\epsilon_3}kl_3)&-\sqrt{\epsilon_3}\sin(\sqrt{\epsilon_3}k l_3)\\
 \frac{\sin(\sqrt{\epsilon_3}k l_3)}{\sqrt{\epsilon_3}}&\cos(\sqrt{\epsilon_3}kl_3)
 \end{bmatrix} \circ \\
 &\begin{bmatrix}
 \cos(kl_2)&-\sin(kl_2)\\
 {\sin(kl_2)}&\cos(kl_2)
 \end{bmatrix} \circ\\
 &\begin{bmatrix}
 \cos(\sqrt{\epsilon_1}kl_1)&-\sqrt{\epsilon_1}\sin(\sqrt{\epsilon_1}kl_1)\\
 \frac{\sin(\sqrt{\epsilon_1}kl_1)}{\sqrt{\epsilon_1}}&\cos(\sqrt{\epsilon_1}kl_1)
 \end{bmatrix}(-i).
 \end{split}
 \end{align}

Similarly to the two-section laser, we expect two fundamentally different types of lasing modes (solutions to \eqref{transfer}). For net local gain in both outer sections, which corresponds to $\text{Im}[\epsilon_{1}]<0$ and $\text{Im}[\epsilon_{3}]<0$, we expect {\em Gain-Neutral-Gain} (GNG) lasing modes. On the other hand, for net local gain in one  outer section and net local absorption in the other outer section, which corresponds to $\text{Im}[\epsilon_{1}]<0$ and $\text{Im}[\epsilon_{3}]>0$ or vice versa, we expect {\em Gain-Neutral-Absorbing} (GNA) lasing modes and {\em Absorbing-Neutral-Gain} (ANG) lasing modes, respectively.  

Using the values in Table \ref{parameters_table}, an example of a GNG 
lasing mode
is shown in Fig.~\ref{riemann_3sec}(a) and  Fig.~\ref{loxo_non_abs}(a). The parameters are chosen to match the green (medium grey) dot in Fig.~\ref{ABS}. The dynamics is governed by the fixed-point structure in each section.  $\mathcal{Z}(z)$ starts out at $-i$ and spirals away from $\mathcal{Z}^F_{11}$ towards $\mathcal{Z}^F_{12}$ on a loxodrome trajectory (blue (dark grey) curve). At $z=l_1$ the vacuum gap causes $\mathcal{Z}(z)$ to follow a circle until $z=z_2$ (orange (light grey)  curve).  In the third section, $\mathcal{Z}(z)$ again follows a loxodrome 
that spirals towards $\mathcal{Z}^F_{32}$ to finish at $\mathcal{Z}(L)=i$  (green (medium grey) curve).  The overall picture in this case is similar to the single-section Fabry Perot case, since both sections 1 and 3 carry net gain.  This is also illustrated in  Fig.~\ref{loxo_non_abs}(c), which shows that the power flow increases in sections 1 and 3. The corresponding electric field intensity is shown in Fig.~\ref{loxo_non_abs}(b). The field intensities in sections 1 and 3 are of comparable magnitude. 

An example of a GNA  lasing mode
is shown in Fig.~\ref{riemann_3sec}(b) and Fig.~\ref{loxo_abs}
for parameters that match the blue (dark grey) dot in Fig.~\ref{ABS}.
Since section 3 is now absorbing,  $\mathcal{Z}(z)$ (green (medium grey) curve) spirals away from $\mathcal{Z}^F_{32}$  before reaching the final point $\mathcal{Z}(L)=i$. As a consequence, the power flow now has a maximum in the inner vacuum section as shown in  Fig.~\ref{loxo_abs}(c).  Fig.~\ref{loxo_abs}(b) indicates that the electric field  intensity in section 3 is significantly smaller than in section 1.

\section{Homogeneously Broadened Media and Cusp Points}
\label{sec:exceptional_points}

Here, we revisit  single-section and three-section lasers from a different perspective. 
Our aim is to reformulate the problem in terms of parameters that correspond to typical physical characteristics of the active-medium, such as gain, or population inversion, and population-induced refractive-index change.
For clarity of exposition, we consider a homogeneously broadened two-level active medium.
For consistency with the single-mode constant-intensity approximation used in Section II, 
we assume constant population inversion in each section. 

To characterise permittivity $\epsilon_j\in\mathbb{C}$ in section $j$ by the active-medium population inversion $N_j$ in section $j$ we use~\cite{haken1985laser,ge2010steady,liertzer2012pump},
\begin{align}\label{z025n}
\epsilon_j&=n_{b,j}^2+\frac{N_j}{\Delta_{j} + i},
\end{align}
where $n_{b,j}\in\mathbb{C}$ is the background refractive index in section $j$, $N_j$ is the population inversion in section $j$, and 
$$
\Delta_{j} =\frac{k-k_{0,j}}{\gamma_{P,j}},
$$
quantifies the population-induced refractive-index change in section $j$; $c\, k_{0,j}$ is the two-level active-medium transition frequency and $c\, \gamma_{P,j}$ is the active-medium polarisation decay in units inverse second.
As a result, the $3n$ independent real parameters listed below Eq.~\eqref{contZ} are replaced by $(6n+1)$ independent parameters:
$\mathrm{Re}(n_{b,1}),\ldots$, $\mathrm{Re}(n_{b,n})$,
$\mathrm{Im}(n_{b,1}),\ldots$, $\mathrm{Im}(n_{b,n})$,
$N_1,\dots,N_n$,
$k_{0,1},\ldots,k_{0,n}$,
$\gamma_{P,1},\ldots,\gamma_{P,n}$,
$k$,
$l_{1},\ldots,l_{n}$.

A significant reduction in the number
of parameters is obtained if we restrict ourselves to particular laser structures, where each section either is a vacuum section, or contains  the same type of an active medium with the possibility of different population inversions in different non-vacuum sections.
Then, the parameters $n_{b,i}$, $k_{0,i}$, and $\gamma_{P,i}$ are the same for all non-vacuum sections, and we denote these global parameters by $n_{b}$, $k_{0}$, and $\gamma_{P}$, respectively.  
As a result, the population-induced refractive-index change $\Delta_j$ is also the same in each non-vacuum section, and  we denote it by $\Delta$. Then,  Eq.~\eqref{z025n} becomes
\begin{align}\label{z028n}
\epsilon_j&=\begin{cases}n_{b}^2+\frac{N_j}{\Delta + i} & \text{for non-vacuum sections}\\
1 & \text{for vacuum sections}
\end{cases}.
\end{align}
Furthermore, we consider $k$ and $\Delta$ to be independent parameters, which further simplifies the problem. In other words,
in a laser with $m$ non-vacuum sections, we have 
$m + n + 3$ real independent parameters:
$\mathrm{Re}(n_b)$, $\mathrm{Im}(n_b)$, $N_1,\ldots,N_m$, $\Delta$,  
$kL$ and the $n-1$ ratios of section lengths $k l_1 : kl_2: \ldots :k l_n$.
In the following, in order to compare our results to the results in \cite{liertzer2012pump},
we allow $kL$, $\Delta$ and the population inversions $N_1,\ldots,N_m$ to vary,
while keeping  the other parameters fixed.

\subsection{Single-Section Laser}

\begin{figure}
	\begin{center}
		\includegraphics[width=\columnwidth]{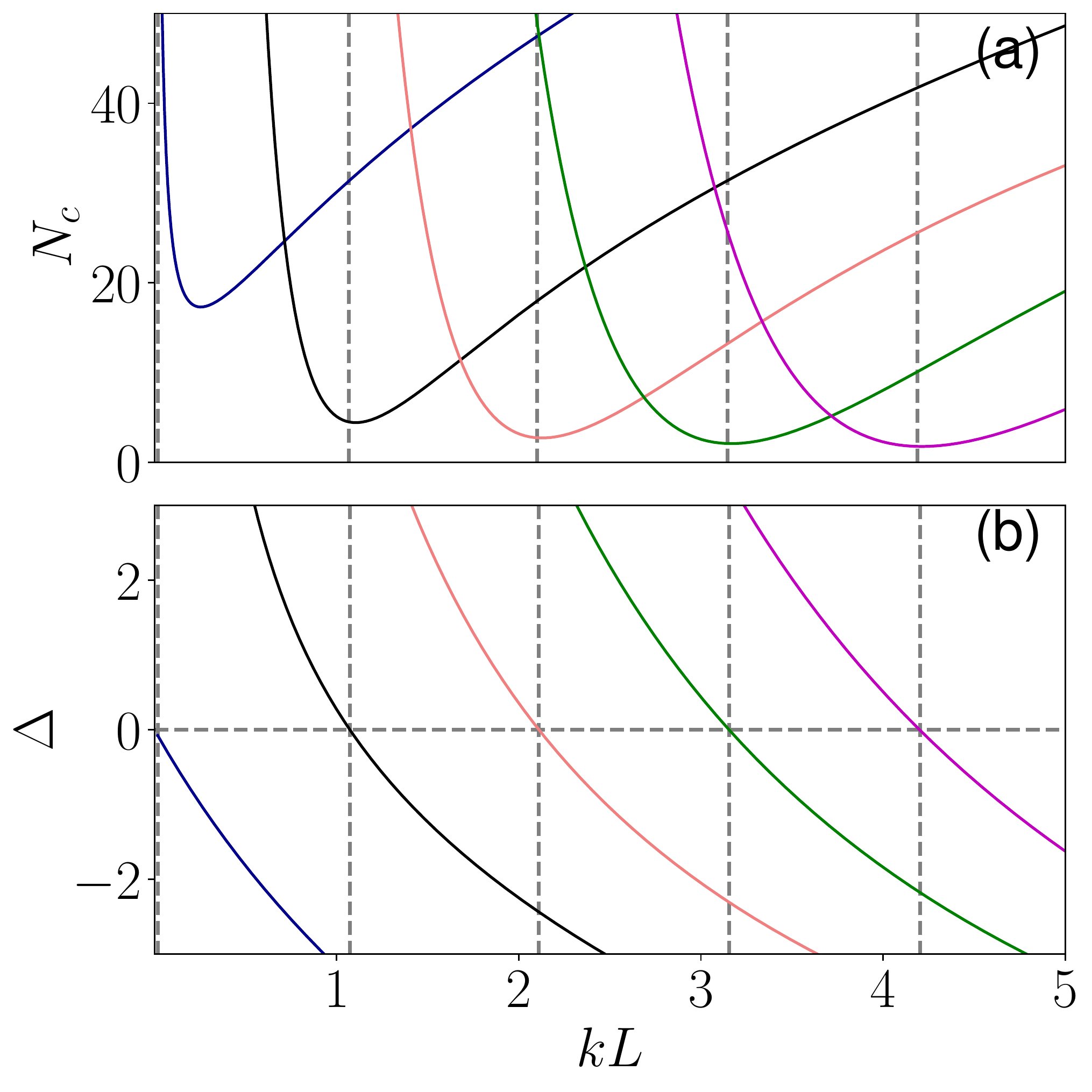}
	\end{center}
	\caption{\label{tep_3} Five solution branches of lasing modes in a single section laser using  equation \eqref{z024} with \eqref{z025} for $n_b=3+0.13i$ projected to the $(kL,N_c)$ plane (a) 
	and  $(kL,\Delta)$ plane (b).}
\end{figure}

In the case of a single-section laser, the permittivity is given by 
\begin{align}\label{z025}
\epsilon_c&=n_b^2+\frac{N_c}{\Delta+i}.
\end{align}

Using \eqref{z025} in the complex equation \eqref{z024}  with a fixed $n_b$ provides two real conditions for the three real parameters $\Delta$, $N_c$ and  $kL$.  The resulting one-dimensional solution branches of lasing modes are shown in Fig.~\ref{tep_3}. Fig.~\ref{tep_3}(a) shows the variation of $N_c$ for the various branches as a function of $kL$.  These solution branches correspond to
the lines shown in Fig.~\ref{tep_2}, and the two figures are related via equation \eqref{z025}.

\begin{figure}
	\begin{center}
		\includegraphics[width=0.9\columnwidth]{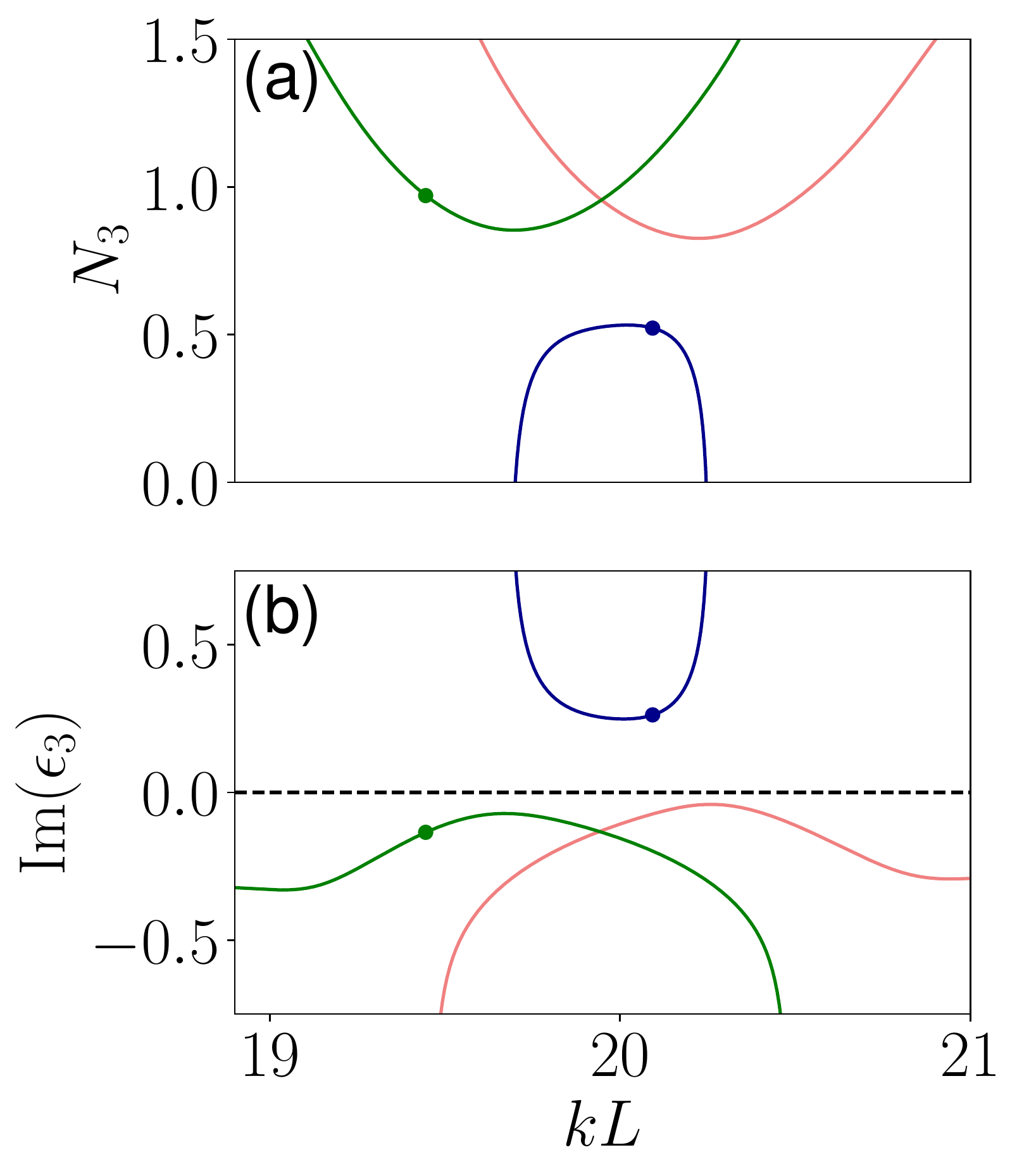}
	\end{center}
\caption{\label{ABS} (a) Solution branches of equation \eqref{transfer} using \eqref{e_1}--\eqref{e_3} for $kl_1:kl_2:kl_3 = 10:1:10$, $n_b=3+0.13i$ and $N_1 = 1.15$ in the $(kL,N_3)$ plane. (b) Corresponding $\text{Im}(\epsilon_3)$ plot using equation \eqref{e_3}. The green (medium grey) and blue (dark grey) dots indicate the values used in Table~\ref{parameters_table}. }
\end{figure}

\begin{figure*}
	\includegraphics[width=2\columnwidth]{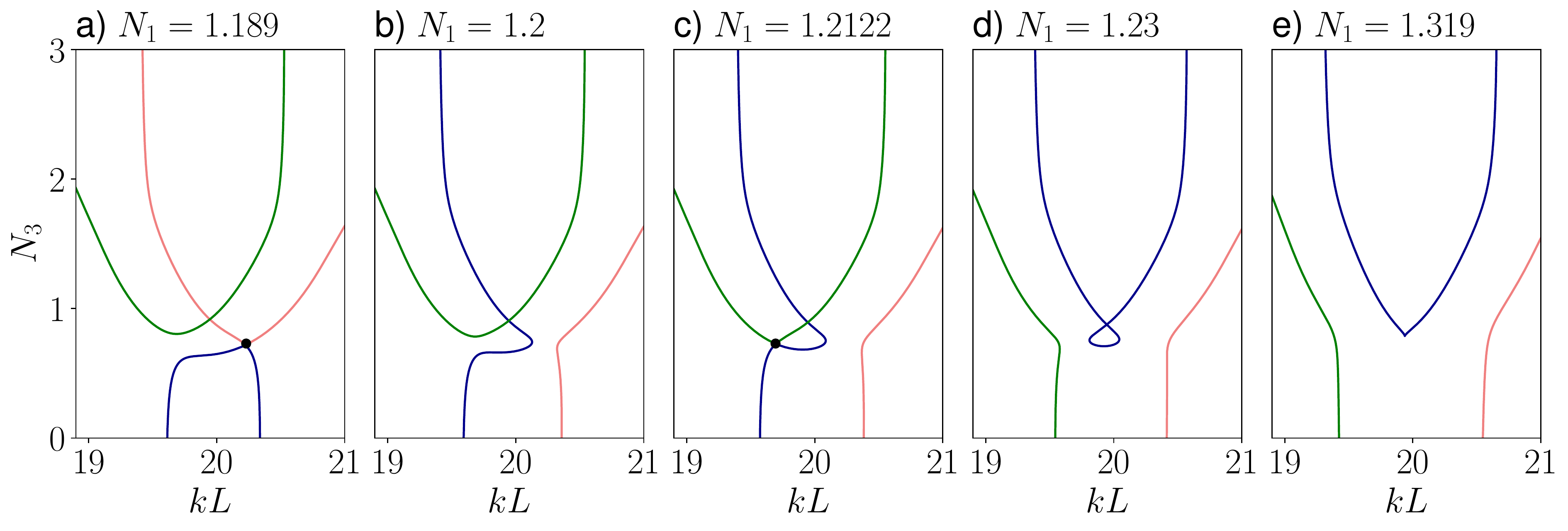}
	\caption{\label{EP_Journey} 
	Solution branches as in Fig.~\ref{ABS}(a) for  different
	values of $N_1$ indicated in each panel. Branch merge points are shown with a black dot.
	}
\end{figure*}
\subsection{Three Section Laser}
Let us now reconsider the three section laser from Section \ref{multi_without_broadening} in the case of homogeneous broadening.
The permittivities in each section are then given by, 
\begin{align}\label{e_1}
\epsilon_1&=n_b^2+\frac{N_1}{\Delta+i},\\ \label{e_2} \epsilon_2&=1\\ \label{e_3}
\epsilon_3&=n_b^2+\frac{N_3}{\Delta+i}.
\end{align}
where $N_1$ and $N_3$ are the population inversion parameters of sections 1 and 3, respectively.  We choose our parameters (see figure captions) to facilitate comparison with \cite{liertzer2012pump}.

Using \eqref{transfer} along with \eqref{e_1}-\eqref{e_3}, we obtain the solution branches of lasing modes as shown in Fig.~\ref{ABS}. We note that the red (light grey) and green (medium grey) branches in Fig.~\ref{ABS}(a) are similar to branches in the single section laser as shown in Fig.~\ref{tep_3}.  Fig.~\ref{ABS}(b) shows that in these cases $\text{Im}(\epsilon_3)$ is negative and therefore section 3 has 
net local 
gain. These branches therefore correspond to GNG 
lasing modes.
However, there also exists a different type of branch, as illustrated by the blue (dark grey) lines in Fig.~\ref{ABS} with an inverted shape and at lower values of $N_3$.  It has \emph{positive} $\text{Im}(\epsilon_3)$  corresponding to 
net local absorption
in section 3
(Fig.~\ref{ABS}(b)) and therefore this lasing mode
is of the GNA type.  This qualitative difference in the branches relates back to our observations in Section V.C where we differentiated solutions with net local gain and 
absorption in section 3. More specifically, the green (medium grey) dots in Fig.~\ref{ABS} correspond to the parameters of Figs.~\ref{riemann_3sec}(a) and \ref{loxo_non_abs}, and the blue (dark grey) dots to those of Figs.~\ref{riemann_3sec}(b) and \ref{loxo_abs}.

It is now interesting to observe, how Fig.~\ref{ABS}(a) changes under variation of a third parameter $N_1$.  This is illustrated in Fig.~\ref{EP_Journey}. These plots reveal a number of interesting phenomena, which we now discuss in detail.  

\begin{figure}
	\includegraphics[width=0.7\columnwidth]{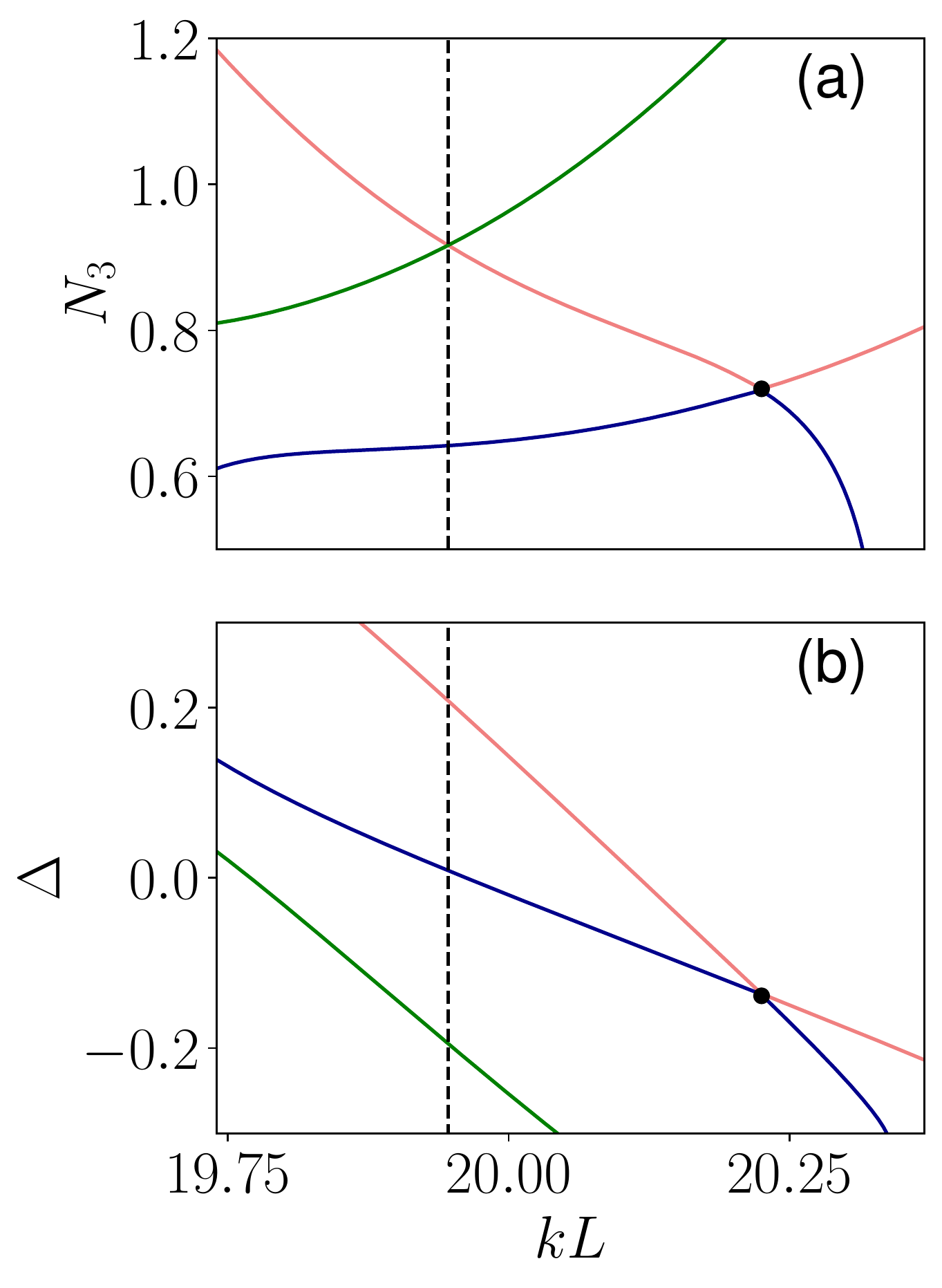}
	\caption{\label{EP_1189} (a) Enlarged version of Fig.~\ref{EP_Journey}(a) close up to the branch merge point (black dot). (b) Corresponding  $(kL,\Delta)$ diagram. 
	}
\end{figure}

To start off, consider the transition from $N_1=1.15$ in Fig.~\ref{ABS}(a) to $N_1=1.189$ in  Fig.~\ref{EP_Journey}(a). We see that the red (light grey) and blue (dark grey) branches meet at 
a special point,
which we call a \emph{branch merge point}.  An enlarged version of the area around this critical point is shown in Fig.~\ref{EP_1189}(a) and a plot of $\Delta$ vs.  $kL$ in (b). Taken together, this demonstrates that the red (light grey) and blue (dark grey) branches 
indeed meet in the three-dimensional $k,N_3, \Delta$ space.  
Note that there is also an apparent crossing of the green (medium grey) and red (light grey) branches at the dotted line in panel (a), which however 
is an artifact of this particular projection: it does not coincide with a crossing in panel (b), and therefore does not correspond to a branch merge point.

Further increase of $N_1$ leads to Fig.~\ref{EP_Journey}(b), where the branches have now a different configuration than in Fig.~\ref{ABS}(a). In particular, both blue (dark grey) and red (light grey) branches now have GNG and GNA solutions and we observe a continuous transition between GNA and GNG 
lasing modes. 
The branches in Fig.~\ref{EP_Journey}(b) correspond  to the threshold boundary discussed in  \cite[Fig.2]{liertzer2012pump}.

As we increase $N_1$ further, we obtain another  branch merge point, shown in Fig.~\ref{EP_Journey}(c). In this case, the green (medium grey) and blue (dark grey) branches merge. After the merge, the blue (dark grey) branch in Fig.~\ref{EP_Journey}(d) develops a peculiar loop.
The green (medium grey) branch now has a continuous transition between GNA and GNG lasing modes.

\begin{figure}
	\includegraphics[width=0.7\columnwidth]{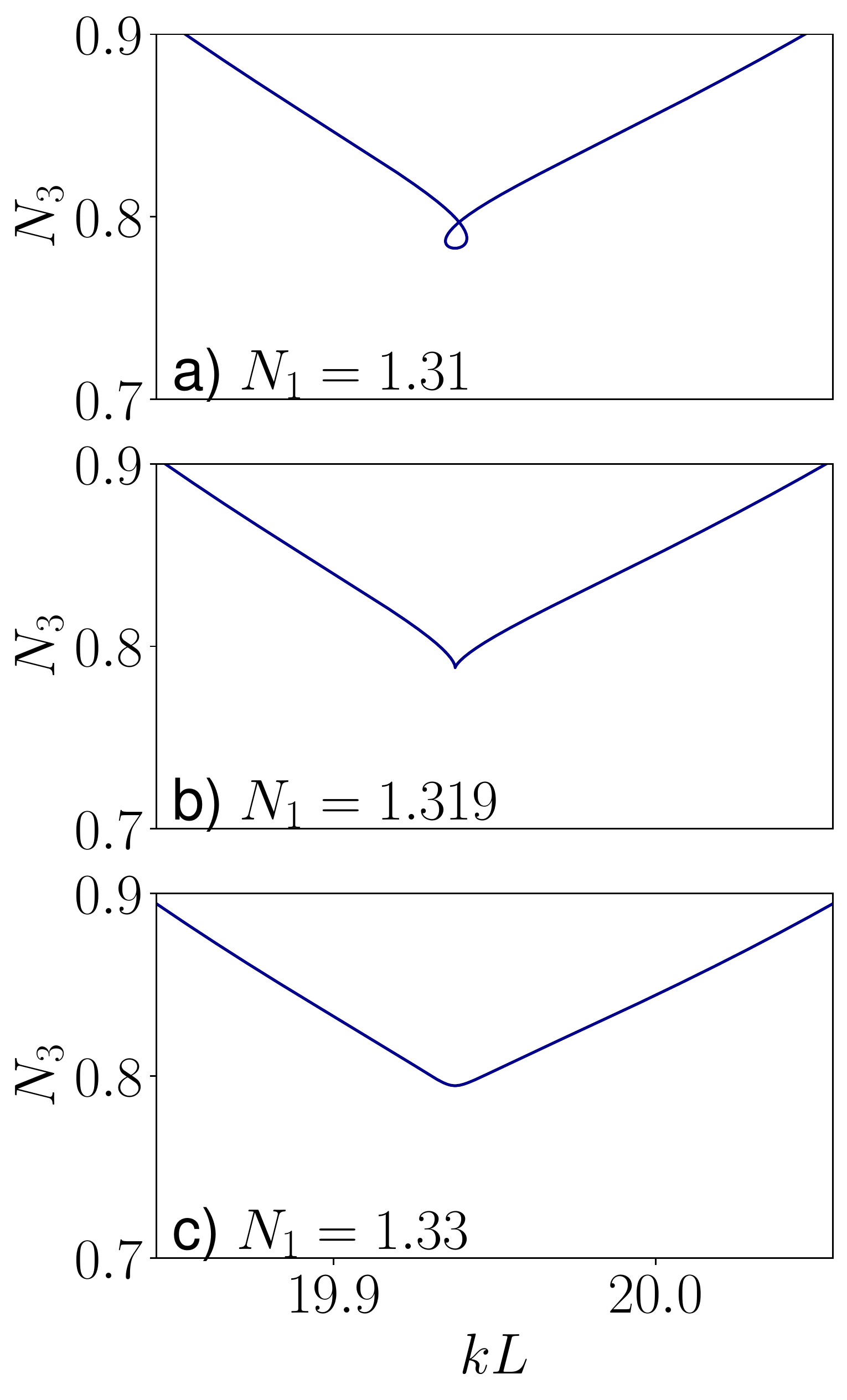}
	\caption{\label{cusp} Magnified solution branch close to  cusp singularity for values near  Fig.~\ref{EP_Journey}(e). }
\end{figure}

Finally, the loop in the blue (dark grey) branch transforms into a {\em cusp singularity} as shown in \ref{EP_Journey}(e).  This is shown in greater detail in Fig.~\ref{cusp}, where we compare the situation slightly before (a), at (b) and after (c) the appearance of the cusp singularity.
We see that at the critical value of $N_1$, the characteristic loop in panel (a) disappears, and the 
blue (dark grey) curve becomes non-smooth with a sharp edge in panel (b). Upon further increase of $N_1$, this edge smooths out as shown in panel (c). This cusp point can be identified with an \emph{exceptional point} at lasing threshold discussed in \cite{liertzer2012pump}. In the formalism from that paper, suitably defined complex ``eigenvalues''  are associated with individual modes, and exceptional points are defined by 
 a degeneracy 
of two such modes.

\subsection{Cusp Point in a Two Section Laser}

\begin{figure}
	\includegraphics[width=\columnwidth]{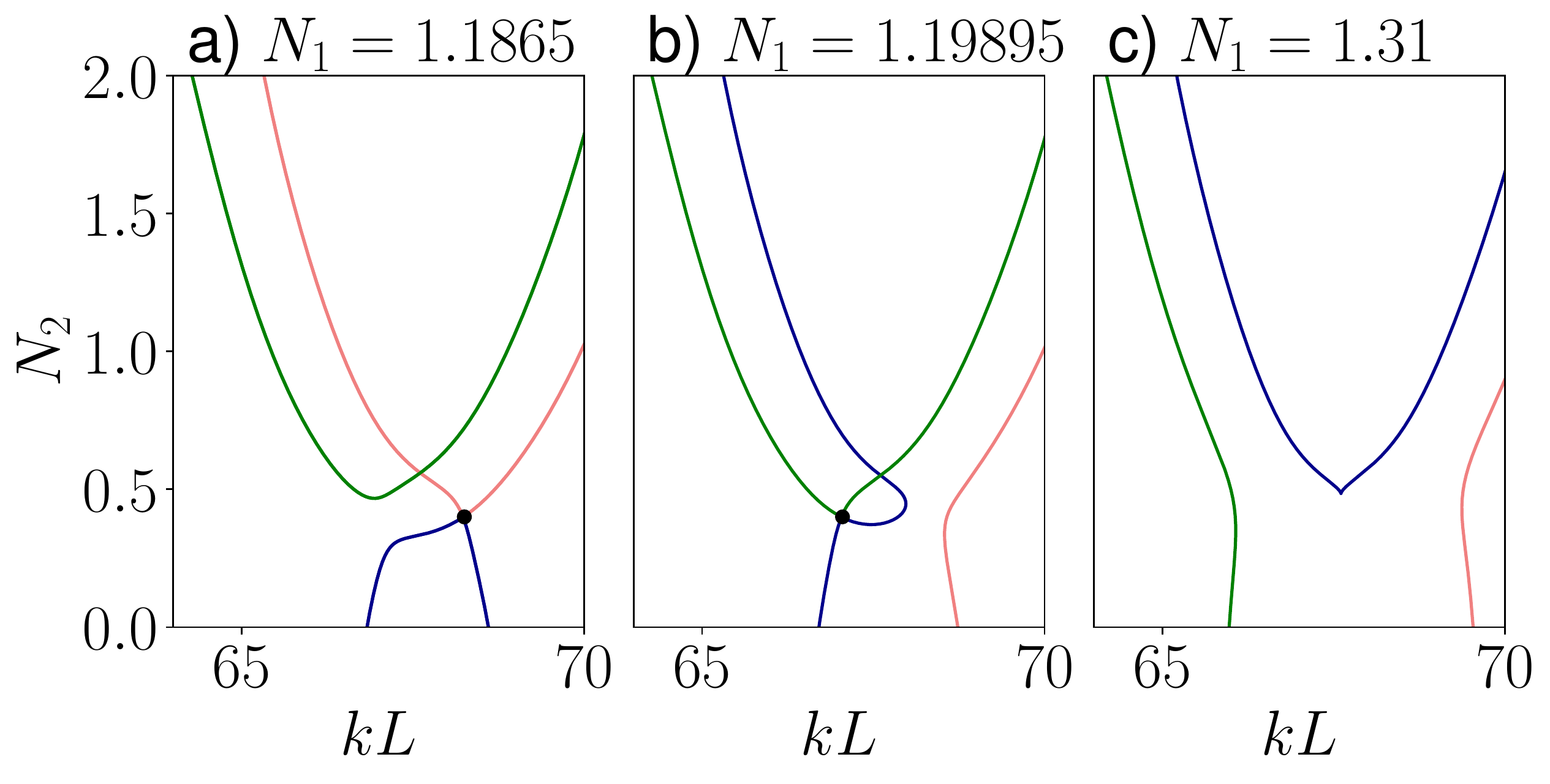}
	\caption{\label{fig:two_section} 
	Solution branches for a two-section laser with	
	 $kl_1:kl_2=1$
	and $n_b=3+0.13i$ for different  values of $N_1$. Branch merge points are shown with a black dot.
	}
\end{figure}

The overall phenomenology of branches described in the previous section for three-section lasers, is also present in the case of two-section lasers, albeit at higher values of $kL$.  To confirm this, Fig.~\ref{fig:two_section} shows the solution branches of lasing modes for a two-section laser with sections of lengths $kl_1:kl_2=1:1$ and homogeneous broadening given by 
\begin{align}\label{two_sec+1}
\epsilon_1&=n_b^2+\frac{N_1}{\Delta+i},\\ \label{two_sec_2}
\epsilon_2&=n_b^2+\frac{N_2}{\Delta+i}.
\end{align}
Fig.~\ref{fig:two_section}(a) and (b) show the merging of two branches analogous to Fig.~\ref{EP_Journey} (a) and (c), respectively. Similarly Fig.~\ref{fig:two_section}(c) represents a cusp point as previously shown in Fig.~\ref{fig:two_section}(e).

\section{Conclusion}
We have investigated the solution space of lasing modes in open boundary multisection lasers with different complex permittivities in each section.  Using suitable mathematical projections, the solutions are conveniently visualized as paths on the Riemann sphere, which start at the point $-i$ and finish at $+i$.  The paths are a continuous concatenation of loxodromes, where each section corresponds to an individual loxodrome.  The mathematical formalism to obtain explicit solutions for the lasing modes involves the use of M\"obius transformations. This method is generally applicable to any number of sections with a different constant permittivity $\epsilon_c$, including piecewise-constant approximations of continuously-varying permittivity profiles $\epsilon(z)$.

The formalism allows us to explore different types of solutions and the connections among them.  In particular, the three section laser exhibits GNG (Gain-Neutral-Gain) and GNA (Gain-Neutral-Absorbing) solutions, which interact in a non-trivial way. In the homogeneously broadened case, we found that two types of critical points exist. The first type are \emph{branch merging points}, where two solution branches merge. This allows for a continuous connection between GNG and GNA solutions. The second type are  \emph{cusp points} which cause the emergence of a characteristic loop in a branch and are analogous to exceptional points at threshold from \cite{liertzer2012pump}.  Very similar behaviour is observed in the two-section laser.

%

\end{document}